\documentclass{ajour}

\usepackage{amsmath}
\usepackage{amsfonts}
\usepackage{epsfig}
\newcommand{\bea}{\begin{eqnarray}}
\newcommand{\eea}{\end{eqnarray}}
\newcommand{\hs}[1]{\hspace*{#1cm}}

\newcommand{\la}{\langle}
\newcommand{\ra}{\rangle}
\newcommand{\half}{{\textstyle \frac{1}{2}}}

\newcommand{\balpha}{\mbox{\boldmath $\alpha$}}

\newcommand{\bbeta}{\mbox{\boldmath $\beta$}}
\newcommand{\bgamma}{\mbox{\boldmath $\gamma$}}
\newcommand{\bepsilon}{\mbox{\boldmath $\epsilon$}}

\newcommand{\bp}{\mbox{\boldmath $p$}}
\newcommand{\bq}{\mbox{\boldmath $q$}}
\newcommand{\bx}{\mbox{\boldmath $x$}}

\newcommand{\br}{\mbox{\boldmath $r$}}
\newcommand{\bA}{\mbox{\boldmath $A$}}

\newcommand{\bS}{\mbox{\boldmath $S$}}

\newcommand{\clf}{\mathcal{F}}
\newcommand{\clh}{\mathcal{H}}

\newcommand{\cll}{\mathcal{L}}
\newcommand{\clp}{\mathcal{P}}
\newcommand{\grad}{\nabla}
\def\tightmaths{                                                                
  \thinmuskip=1.5mu                                                             
  \medmuskip=2mu plus 1mu minus 2mu                                             
  \thickmuskip=2.5mu plus 2.5mu
}                         
                                  
\tightmaths

\begin{document}

%


\authorrunninghead{C. E. Dolby and S. F. Gull}
\titlerunninghead{Fermionic QFT In Gravitational Backgrounds}





\title{Radar Time and a State-Space Based Approach To Quantum Field Theory 
In Gravitational and Electromagnetic Backgrounds}


\authors{Carl E. Dolby and Stephen F. Gull}
\affil{Astrophysics Group, Cavendish Laboratory, Madingley Road, Cambridge
CB3 0HE, U.K.}

\email{c.dolby@mrao.cam.ac.uk}

\abstract{In a recent paper \cite{Me1} a new initial value formulation of 
fermionic QFT was presented that is applicable to an arbitrary observer in any 
electromagnetic background. This 
approach suggests a consistent particle interpretation at all times, with the 
concept of `radar time' used to 
generalise this interpretation to an arbitrarily moving observer. In the present  
paper we extend this formalism to allow for gravitational backgrounds. The 
observer-dependent particle interpretation generalises Gibbons' 
definition \cite{Gibb2} to non-stationary spacetimes. This allows any observer 
to be considered, providing a particle interpretation that depends {\it only} 
on the observer's motion and the background, not on any choice of 
coordinates or gauge, or on details of their particle detector. 
Consistency with known results is demonstrated for the cases of Rindler space 
and deSitter space. Radar time is also considered for an arbitrarily moving 
observer in an arbitrary 1+1 dimensional spacetime, and for a comoving observer 
in a 3+1 dimensional FRW universe with arbitrary scale factor $a(t)$. Finite 
volume measurements and their fluctuations are also discussed, allowing one to 
say with definable precision where and when the particles are observed.}

\keywords{particle creation, fermion, observer, Slater determinant, radar time.}

\begin{article}

\section{INTRODUCTION}

Our recent initial value formulation of fermionic QFT \cite{Me1} emphasised the 
states of the system, described in 
terms of Slater
determinants of Dirac states. The vacuum was defined as
the Slater determinant of a basis for the span of the
negative spectrum of the `first quantized' Hamiltonian, thus providing  
a concrete manifestation of the Dirac Sea. As well as generating 
simple derivations of the general S-Matrix element and 
expectation value in the theory, the approach suggested 
a consistent particle interpretation at all times. In the present 
paper we extend this work to encompass gravitational backgrounds.

In common with the Canonical, Tunnelling, and Wave-functional approaches,
our analysis depends explicitly on the choice of `in' and `out'
particle/antiparticle decomposition, and on the corresponding categorisation
of in and out modes. Various categorisation schemes have been
proposed, based on asymptotic or adiabatic properties of
solutions, or on the diagonalisation of a suitable Hamiltonian.
However, most schemes depend on a choice of coordinates or of gauge~\cite{SPad},
and the relation of these choices to the motion of an observer or the behaviour of a 
particle detector is often ambiguous~\cite{Sr1}. Also, those schemes based on
asymptotic or adiabatic approximations generate accurate
predictions only at asymptotically late times or in sufficiently
weak backgrounds. Particle detectors provide a
more operational particle concept, but their predictions are not
always proportional to the number of particles
present~\cite{Sr1,Sr2} even when the detector is inertial.

    The categorisation scheme proposed in \cite{Me1} and developed here provides
a consistent particle interpretation at all times without requiring any
asymptotic conditions on the in and out states. It consistently combines the conventional
`Bogoliubov coefficient' and `tunnelling amplitude' methods, resolving
gauge inconsistencies that trouble each of these~\cite{SPad}. Also,
by utilising the concept
of radar time, the present definition naturally incorporates the motion of the
observer (detector), providing a definition which depends only on the observer's
motion and on the background, and not on the choice of coordinates or gauge. Since 
it is applicable at all times, we can state not only how many 
particles are created, but also when they were created, and how they 
behaved after their creation (as in the application of \cite{Me1} to 
spatially uniform electric fields \cite{Me3}). By considering finite-volume operators 
with controllable fluctuations, we can further specify, with 
definable precision, where the particles are created.

In Section 2 we specify the representation of states used, 
and their evolution, and present formulae for the general S-Matrix element and 
the expectation value of the theory. The particle interpretation is described 
in Section 3, specifying the states of Section 2 in terms of 
their particle content. Finite volume measurements are also discussed in 
this Section, along with their fluctuations. For concreteness we suppose that the 
spacetime is globally hyperbolic, and that the observer-dependent foliation is 
Cauchy. Strictly, this excludes situations with particle horizons, but we present 
a simple example in Section 4 which shows that the formalism is 
still well suited to the 
treatment of horizons. We consider massive and massless Dirac fermions in 1+1 dimensional 
flat space as seen by a uniformly accelarating observer, and demonstrate  
consistency with known techniques, by rederiving the well-known thermal 
distribution of Rindler particles. We consider 
the spatial distribution of these Rindler particles, and comment briefly on fluctuations.

	In Section 5 we present examples of radar time, treating an 
arbitrary observer in an arbitrary 1+1 dimensional spacetime, and also a comoving 
observer in an FRW universe with arbitrary scale factor $a(t)$. DeSitter space 
and the Milne universe are described in more detail. We conclude with a brief 
discussion in Section 6. An appendix describes the connection 
between projection operators and 2-point functions, and emphasises the role of 
the negative energy Wightman function as the ``Dirac density matrix of 
the Dirac Sea''.

\section{THE STATE SPACE}

\subsection{Preliminaries}

The Lagrangian density for the Dirac equation in gravitational and 
electromagnetic backgrounds is \cite{BD,GMR}

\begin{equation} \cll = \Re[e(x) \bar{\psi}(x)(i \gamma^{\mu}(x) \grad_{\mu} - m) 
\psi(x)] \label{eq:diss7.1}\end{equation}
	where $\gamma^{\mu}(x) = e^{\mu}_{a}(x) \bar{\gamma}^a$ 
satisfies $ \{ \gamma^{\mu}(x), \gamma^{\nu}(x) \} = 2 g^{\mu \nu}(x) $, 
	$\bar{\gamma}^a$ are a representation of the normal flat
	space Dirac matrices, $\bar{\psi} \equiv \psi^{\dagger}
	\bar{\gamma}^0$, $e^{\mu}_{a}(x)$ is the {\it Vierbein}, and 
$e(x) \equiv \det(e^a_{\mu}(x)) = \sqrt{- \det(g_{\mu \nu}(x))}$. The 
covariant derivative $\grad_{\mu}$ acts on spinors according to:

\begin{equation} \grad_{\mu} \psi(x) = (\partial_{\mu}  + \Gamma_{\mu} + 
i e A_{\mu}) \psi(x) \notag \end{equation}
where $\Gamma_{\mu} = \frac{1}{4} \gamma_{\nu} \grad_{\mu} 
\gamma^{\nu}$, $A_{\mu}$ is the electromagnetic potential, and $e$ 
is the charge of the fermion ($e < 0$ for electrons). $\Gamma_{\mu}$ 
is often written \cite{Kak} in terms of the `connection field' 
$\omega_{\mu}^{a b}(x)$ as $\Gamma_{\mu} = - \frac{i}{4} \omega_{\mu}^{a b} 
\sigma_{a b}$ where $\sigma_{a b} = \frac{i}{2} [\bar{\gamma}_a,\bar{\gamma}_b ]$.

	This Lagrangian gives rise to the governing equation:
\begin{equation} (i \gamma^{\mu}(x) \grad_{\mu} - m) \psi(x) = 0 
\label{eq:diss7.2}\end{equation}
and the energy-momentum tensor: 

\begin{equation} T_{\mu \nu}(\psi) = \Re [ i \bar{\psi} \gamma_{(\mu} 
\grad_{\nu)} \psi ] - \frac{g_{\mu \nu}}{e(x)} \cll \label{eq:Tgrav} \end{equation}
The inner product 
$\la \psi | \phi \ra_{\Sigma}$ on the
	spacelike Cauchy surface $\Sigma$ is given by:
\begin{equation} \la \psi | \phi \ra_{\Sigma} = \int e(x) \bar{\psi}(x) 
\gamma^{\mu}(x) \phi(x) d \Sigma_{\mu} \label{eq:diss7.5}\end{equation}
	and is independent of $\Sigma$ by virtue of
	(\ref{eq:diss7.2}).

The `first quantized' state space $\clh(\Sigma)$ on some spacelike hypersurface
$\Sigma$ can be defined as the space of all (finite norm) spinor valued 
functions of $x|_{\Sigma}$ (the projection of $x^{\mu}$ onto 
the hypersurface $\Sigma$). We restrict our attention to 
hypersurfaces $\Sigma$ that are Cauchy, so that the various $\clh(\Sigma)$ 
are all unitarily equivalent, and can simply be denoted $\clh$. We will 
denote a first quantized state on $\Sigma$ by $\psi(x|_{\Sigma})$ or  
$|\psi_{\Sigma} \ra$ or, where no ambiguity is possible, simply $\psi$. (There 
will be little need to distinguish between a state and its coordinate representation.)

\subsection{The Full Fock Space Over $\clh$}

The {\it antisymmetric Fock Hilbert space} over the complex Hilbert
space $\clh$ is denoted $\clf_{\wedge}(\clh)$ and is defined \cite{Ott} in
terms of the {\it antisymmetric Tensor Algebra} over $\clh$. It is 
a natural and familiar construction by which a
 quantum theory of fermions can be formulated. This construction is described 
in \cite{Me1}, and as a preliminary we outline it here. Let $\clh$
be the Hilbert space in the previous Section, with inner product denoted by 
$\la \hs{.2}| \hs{.2} \ra$. Let $\otimes^n \clh$ denote the direct product of 
$n$ copies of
$\clh$, and let $\wedge^n \clh$ denote the restriction of $\otimes^n
\clh$ to those states which are completely antisymmetric under
changes in the order of the elements $|\psi\ra \in \clh$ from which the state is
constructed. Given $|\psi_1\ra |\psi_2\ra \dots |\psi_n \ra 
\in
\otimes^n \clh$, we can define $| \psi_1 \wedge \psi_2 \wedge \dots 
\wedge \psi_n \ra \in \wedge^n \clh$ by:
\begin{equation}| \psi_1 \wedge \psi_2 \wedge \dots 
\wedge \psi_n \ra \equiv 
\frac{1}{\sqrt{n!}} \sum_{\sigma} {\rm sign}(\sigma) |\psi_{\sigma(1)}\ra 
|\psi_{\sigma(2)}\ra \dots |\psi_{\sigma(n)}\ra \label{eq:oplus}\end{equation}
	where $\{ \sigma(i), i=1,\dots ,n \}$ is a permutation of 
$\{ 1 \dots n \}$. This is simply the Slater determinant of the states 
$|\psi_1\ra \dots |\psi_n\ra$. The antisymmetric Fock Hilbert space 
$\clf_{\wedge}(\clh)$ is now given by:
\begin{equation} \clf_{\wedge}(\clh) = \oplus_{n=0}^{\infty} \wedge^n 
\clh \notag \end{equation}
	where $\wedge^0 \clh \equiv \mathbb{C}$ and $\wedge^1 \clh
	\equiv \clh$. States which lie entirely within $\wedge^r \clh$
	for some $r$ are said to
	be of {\it grade} r. 

A useful operation on $\clf_{\wedge}(\clh)$ is the
`inner derivative' $i_{\psi} : \wedge^n \clh
\rightarrow \wedge^{n-1} \clh$ (named by analogy with differential
geometry). This is defined by:
\begin{equation} i_{\psi} | \phi_1 \wedge \dots \wedge \phi_n \ra \equiv  
 \sum_i (-)^{i+1} \la \psi | \phi_i \ra | \phi_1 \wedge
 \dots \wedge \check{\phi}_i \wedge \dots \wedge \phi_n \ra  
\label{eq:dot}\end{equation} 
	 where the check over $\phi_i$ signifies that this state is omitted 
from the product. The relation $i_{\psi} : \clf_{\wedge}(\clh)
\rightarrow \clf_{\wedge}(\clh)$ is obtained from (\ref{eq:dot}) by imposing 
linearity, together with the convention $i_{\psi} \lambda = 0$ for 
$\lambda \in \wedge^0 \clh$. It is clear that $i_{\psi} (i_{\psi} | F \ra) = 0$ for all 
$| F \ra \in \clf_{\wedge}(\clh)$, and that:
\begin{equation} i_{\psi} (\phi \wedge |F \ra) = \la \psi | \phi \ra |F \ra -
\phi \wedge (i_{\psi} | F \ra) \label{eq:CARprim} \end{equation}
The operation $i_{\psi}$ is denoted $a(\psi)$ by Ottesen \cite{Ott}, and plays the
role of an annihilation operator. Here $i_{\psi}$ will play a
similar, although not identical role.

	Finally, the inner product on 
$\clf_{\wedge}(\clh)$ is given by: 
\begin{equation} \la \psi_1 \wedge \dots \wedge \psi_n | 
\phi_1 \wedge \dots \wedge \phi_m \ra = \delta_{n m} 
\det [ \la \psi_i | \phi_j \ra ] \label{eq:inprod} \end{equation}
	where $\la \psi_i | \phi_j \ra$ refers to the inner 
	product on $\clh$. (For states $\lambda,\mu \in \wedge^0 \clh$ define 
$\la \lambda | \mu \ra = \bar{\lambda} \mu$ and $\la \lambda | F_n \ra = 0$ for any 
state $|F_n\ra$ of grade $n > 0$.) This definition agrees with the inner
	product defined in terms of Slater determinants. 
Although we use the notation $\la \hs{.2} | \hs{.2} \ra$ to refer to both the inner product 
on $\clh$ and the inner product on $\clf_{\wedge}(\clh)$, 
it will always be clear from the context which is involved.

\subsection{Operators on Fock Space}

Let $\hat{A}_1:\clh \rightarrow \clh$ be an operator on the space of
Dirac states.  We wish to construct from it an operator which can act
on all of state space. There are two useful ways of doing this: 
{\it Hermitian extension} $\hat{A}_H : \clf_{\wedge}(\clh)
\rightarrow \clf_{\wedge}(\clh)$, and {\it Unitary extension}
$\hat{A}_U: \clf_{\wedge}(\clh) \rightarrow \clf_{\wedge}(\clh)$ (outlined also in
Ottesen \cite{Ott}). These are respectively defined by:

\begin{align} \hat{A}_H | \psi_1 \wedge \psi_2 \wedge \dots 
\wedge \psi_n \ra & \equiv  \sum_{i=1}^{n} | \psi_1 \wedge \dots (\hat{A}_1 \psi_i) \wedge 
\psi_{i+1} \dots \wedge \psi_n \ra \label{eq:SSB8} \\
 \hat{A}_U | \psi_1 \wedge \psi_2 \wedge \dots 
\wedge \psi_n \ra & \equiv | (\hat{A}_1\psi_1) \wedge (\hat{A}_1\psi_2) 
\wedge \dots \wedge (\hat{A}_1\psi_n) \ra \label{eq:SSB9}\end{align}

If $\hat{A}_1$ is (anti)hermitian
with respect to the inner product (\ref{eq:diss7.5}) on $\clh$, then 
$\hat{A}_H$ is (anti)hermitian
with respect to the inner product (\ref{eq:inprod}) on 
$\clf_{\wedge}(\clh)$. If $\hat{A}_1$ is unitary,
then so is $\hat{A}_U$. Also $(e^{\hat{A}_1})_U =
e^{\hat{A}_H}$, so that if $\hat{U}_1 = e^{\hat{A}_1}$ on $\clh$ 
then $\hat{U}_U = e^{\hat{A}_H}$ on
$\clf_{\wedge}(\clh)$.

\subsubsection*{Some Simple Properties}

\begin{enumerate}

\item $(\hat{A} + \hat{B})_H = \hat{A}_H + \hat{B}_H$, $[ 
\hat{A}_H,\hat{B}_H ] = [ \hat{A} , \hat{B} ]_H$ and $(\hat{A}
\hat{B})_U = \hat{A}_U \hat{B}_U$.

\item $[ \hat{A}_H , \psi \wedge ] = (\hat{A}_1 \psi) $ and 
$[ \hat{A}_H , i_{\psi} ] = - i_{\hat{A}^{\dagger}_1 \psi}$ 

\item If $ \psi_1, \psi_2, \dots \psi_n $ are all 
eigenstates of $\hat{A}_1$
with eigenvalues $\lambda_1, \dots \lambda_n$, then $| \psi_1 \wedge 
\psi_2 \dots \wedge \psi_n \ra $ is an eigenstate of $\hat{A}_H$ with 
eigenvalue $\sum_{i = 1}^{n} \lambda_i$.

\item If  $ \psi_1, \psi_2, \dots \psi_n $ are orthonormal 
and $| F \ra \equiv | \psi_1 \wedge \psi_2 \dots \wedge \psi_n \ra$
then 
\begin{align} \la F | \hat{A}_H | F \ra
 & = \sum_{i = 1}^{n} \la \psi_i | \hat{A}_1 | \psi_i \ra \label{eq:prop3a}\\
\la F |(\hat{A}_H)^2 | F \ra 
& = \sum_{i = 1}^{n} \la \psi_i | \hat{A}_1^2 | \psi_i \ra  + 
 2 \sum_{i < j} ( \la \psi_i | \hat{A}_1 | \psi_i \ra \la \psi_j | \hat{A}_1 
| \psi_j \ra \label{eq:prop3b} \\ 
& \hs{4} - \la \psi_i | \hat{A}_1 | \psi_j \ra \la \psi_j | \hat{A}_1 
| \psi_i \ra) \notag \\
 \la F |(\hat{A}_{H})^2 | F \ra & - (\la F | \hat{A}_{H} | F \ra)^2 = 
\sum_i  \la \psi_i | \hat{A}_1^2 | \psi_i \ra  -  
\sum_{i,j} | \la \psi_i |\hat{A}_1 |\psi_j \ra |^2 \label{eq:prop3c} \end{align}

\end{enumerate}

\subsection{Evolution of States}
 
  Given two Cauchy surfaces $\Sigma_1$ and $\Sigma_0$, we can define 
the evolution operator $\hat{U}_1(\Sigma_1,\Sigma_0)$ on $\clh$ by

\begin{equation} \hat{U}_1(\Sigma_1,\Sigma_0) |\psi_{\Sigma_0} \ra 
\equiv |\psi_{\Sigma_0}(\Sigma_1)\ra 
\label{eq:SSB25}\end{equation}
 where $|\psi_{\Sigma_0}\ra$ represents some chosen initial conditions
 $\psi(x|_{\Sigma_0})$ on $\Sigma_0$, and $|\psi_{\Sigma_0}(\Sigma_1)\ra$ 
represents the corresponding solution $\psi(x)$ of the Dirac equation, 
expressed on the hypersurface $\Sigma_1$.

	We consider only QFT in an (external) gravitational or electromagnetic background, 
so that we ignore direct particle-particle interactions and work within 
the `zeroth order Hartree-Fock' approximation. This assumes that the 
evolution operator on $\clf_{\wedge}(\clh)$ is just the unitary extension of 
the evolution operator on $\clh$. The action of $\hat{U}(\Sigma_1,\Sigma_0)$ 
is now given by:

\begin{equation} \hat{U}(\Sigma_1,\Sigma_0) | \psi_{1,\Sigma_0} \wedge 
\dots \wedge \psi_{n,\Sigma_0} \ra = | \psi_{1,\Sigma_0}(\Sigma_1) \wedge 
\dots \wedge \psi_{n,\Sigma_0}(\Sigma_1) \ra \label{eq:evol} \end{equation}
The multiparticle solution is simply the Slater 
determinant of the appropriate `first quantized' solutions. This 
construction preserves grade, and implies that the unitarity of 
$\hat{U}(\Sigma_1,\Sigma_0)$ follows immediately from the 
unitarity of the first quantized Dirac equation.

We have now set up a state space, an evolution equation (\ref{eq:evol}) and a 
conserved inner product (\ref{eq:inprod}). These are all we need to calculate 
arbitrary S-Matrix elements (from (\ref{eq:evol}) 
and (\ref{eq:inprod})), arbitrary expectation values (from (\ref{eq:prop3a})), 
and even fluctuations in these expectation 
values (from (\ref{eq:prop3c})). However,  
the theory is not invested with physical meaning 
until the states of the system can be specified in terms of their physical 
properties. For this purpose a particle interpretation is invaluable. 
We now introduce an observer-dependent 
particle interpretation. This will rely on the observer-dependent 
foliation of spacetime provided by Bondi's Radar Time \cite{Bondi,Bohm,Me3}.

\section{OBSERVER DEPENDENT PARTICLE INTERPRETATION}

\subsection{Bondi's Radar Time}

Consider an observer travelling on path $\gamma: x^{\mu} = x^{\mu}(\tau)$ 
with proper time $\tau$, and define:
\begin{align}
\tau^{+}(x) & \equiv \mbox{ (earliest possible) proper time at which a 
null geodesic leaving} \notag \\
& \hs{2} \mbox{ point $x$ could intercept $\gamma$. } \notag \\
\tau^{-}(x) & \equiv \mbox{ (latest possible) proper time at which a null 
geodesic could} \notag \\
& \hs{2} \mbox{ leave $\gamma$, and still reach point $x$. } \notag \\
\tau(x) & \equiv \half (\tau^{+}(x) + \tau^{-}(x)) \hs{1} = 
\mbox{ `radar time'.} \notag \\
\rho(x) & \equiv \half (\tau^{+}(x) - \tau^{-}(x)) \hs{1} = 
\mbox{ `radar distance'.} \notag \\
\Sigma_{\tau_0} & \equiv \{x: \tau(x) = \tau_0 \} = \mbox{ observer's `hypersurface 
of simultaneity at time $\tau_0$'. } \notag \end{align}

\begin{figure}[h]
\vspace{-.3cm}
\center{\epsfig{figure=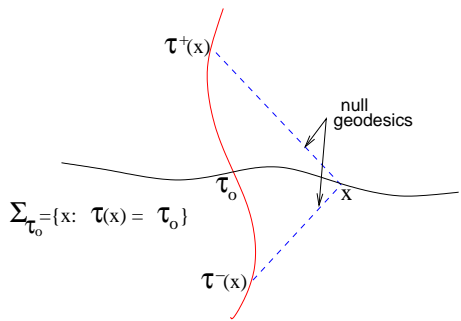 , width=6cm}}
\caption{{\footnotesize Schematic of the definition of `radar time' $\tau(x)$.}}
\vspace{-.3cm}
\end{figure}

This is a simple generalisation of the definition made 
popular by Bondi in his work on 
special relativity and {\it k}-calculus \cite{Bondi,Bohm,Dinverno}. By 
construction radar time is independent of the choice of coordinates (since no 
coordinates need be introduced in defining it) and it depends only on the 
motion of the observer.   
 It agrees with proper time on the 
observer's path, and is invariant under `time-reversal' - that is, 
under reversal of the sign of the observer's proper time.

	It is clear from the definition of radar time that $\Sigma_{\tau_1}$ lies to the 
future of $\Sigma_{\tau_0}$ (for $\tau_1 > \tau_0$) except at the observer's 
particle horizon (if one exists), on which the various $\Sigma_{\tau}$ 
converge. When horizons are present, the domain of $\tau(x)$ is no longer all 
of spacetime - only that part with which the observer can send and receive signals.

	Define now the `time-translation' vector field: 
\begin{equation} k_{\mu}(x) \equiv \frac{\frac{\partial \tau}{\partial 
x^{\mu}}}{g^{\sigma \nu} \frac{\partial \tau}{\partial x^{\sigma}} 
\frac{\partial \tau}{\partial x^{\nu}}} \label{eq:time} \end{equation}
	This represents the perpendicular distance between neighbouring hypersurfaces of 
simultaneity, since it is normal to these hypersurfaces and it satisfies 
$k^{\mu}(x)\frac{\partial \tau}{\partial x^{\mu}} = 1$. Now use the identity 
$i \hs{.1}|\hs{-.15}k \gamma^{\mu} \grad_{\mu} = i k^{\mu} \grad_{\mu} + 
\sigma^{\mu \nu} k_{\mu} \grad_{\nu}$  (where $\hs{.1}|\hs{-.15}k \equiv k_{\mu} \gamma^{\mu}$ 
and $\sigma^{\mu \nu} \equiv \frac{i}{2} [ \gamma^{\mu} , 
\gamma^{\nu}] = \sigma^{a b} e^{\mu}_{a} e^{\nu}_{b}$) to write the Dirac equation as:
\begin{equation} i k^{\mu} \grad_{\mu} \psi = \hat{H}_{{\rm nh}}(\tau) \psi \equiv - \sigma^{\mu \nu} k_{\mu} 
\grad_{\nu} \psi + m \hs{.1}|\hs{-.15}k \psi \label{eq:dissHam2}\end{equation}

Here $\hat{H}_{{\rm nh}}(\tau)$ is not in general Hermitian! (hence the 
subscript nh). At first sight this seems to 
disagree with unitarity on $\clh$. However there is no  
inconsistency, because the inner product now depends explicitly on 
$\tau$ (via the volume element on $\Sigma_{\tau}$), and since   
(\ref{eq:dissHam2}) is no longer of the form $i \frac{d}{d t} | \psi(t) \ra = \hat{H}_1(t) 
| \psi(t) \ra$, invalidating the standard equivalence proof of unitary evolution 
 and a Hermitian Hamiltonian. 

	To investigate the relation between $\hat{H}_{{\rm nh}}(\tau_0)$ 
and the energy momentum tensor, define:

\begin{equation} H_{\tau_0}(\psi) \equiv \int_{\Sigma_{\tau_0}} T_{\mu \nu}(\psi(x)) 
k^{\mu} {\rm d} 
\Sigma^{\nu} \label{eq:Hsig} \end{equation}
	Substitution of (\ref{eq:Tgrav}) into this gives:

\begin{gather} H_{\tau_0}(\psi) = \Re [ \la \psi | \hat{H}_{{\rm nh}}(\tau_0) | \psi 
\ra_{\Sigma_{\tau_0}} ] = \la \psi | \hat{H}_1(\tau_0) | \psi \ra_{\Sigma_{\tau_0}} 
\label{eq:Hsig2} \\
 \mbox{ where } \hat{H}_1(\tau_0) \equiv \half \{ \hat{H}_{{\rm nh}}(\tau_0) +  
\hat{H}^{\dagger}_{{\rm nh}}(\tau_0) \} \notag \end{gather}
We can define the projection operators 
$\hat{P}^{\pm}_{\tau_0}: \clh \rightarrow \clh^{\pm}(\tau_0)$, 
and hence the spaces 
$\clh^{\pm}(\tau_0)$, by requiring that $\hat{P}^{\pm}_{\tau_0}$ are 
orthogonal projections satisfying:
\begin{equation} H_{\tau_0}(\hat{P}^+_{\tau_0} \psi) 
\geq H_{\tau_0}(\psi) \geq 
H_{\tau_0}(\hat{P}^-_{\tau_0} \psi) \label{eq:Hdef} \end{equation}
for all $\psi \in \clh$, as in \cite{Me2}. 
This definition depends only on the background and the motion of the observer, 
 and not on the choice of coordinates 
or gauge. It is equivalent to defining:
\begin{align} \clh^+(\tau_0) & \mbox{ as the span of the positive spectrum of } 
	\hat{H}_1(\tau_0) \notag \\
\clh^-(\tau_0) & \mbox{ as the span of the negative spectrum of }
	\hat{H}_1(\tau_0) \notag \end{align}

Here $\clh^+(\tau_0)$ is the set of all positive energy states, and 
$\clh^-(\tau_0)$ is the set of all negative energy states as defined 
on $\Sigma_{\tau_0}$. This definition generalises Gibbons' approach 
\cite{Gibb2} to arbitrary observers and non-stationary spacetimes.

In the Canonical approach to background QFT, this 
split of $\clh$ into positive/negative energy modes can 
be achieved by Hamiltonian diagonalisation of the second quantized 
Hamiltonian that arises by substituting the field operator 
$\hat{\psi}(x)$ into expression (\ref{eq:Hsig}) 
for $H_{\tau_0}(\psi)$. Hamiltonian diagonalisation has been
criticised~\cite{Full2} for its  reliance on the choice of a second quantized 
Hamiltonian which depends on an apparently
arbitrary choice of hypersurface and time-translation vector field $k^{\mu}$. Here this 
arbitrariness has been resolved (and in \cite{Me1,mythesis}) by 
specifying the hypersurface $\Sigma_{\tau_0}$ and the vector field 
$k^{\mu}$ in terms of the worldline of the observer (or particle
detector).

	We have assumed here that $\hat{H}_1(\tau)$ has no zero energy 
eigenstates (for any $\tau$). Even for inertial observers, there exist 
topologically non-trivial backgrounds for which zero energy 
eigenstates exist, leading to the existence of fractional charge; see e.g.  
\cite{Jackiw}. Although such situations 
are straightforward to describe within the present approach, we will not 
discuss them further here. When there is a particle horizon, 
so that the observer's hypersurfaces are not Cauchy, zero-energy 
eigenstates are plentiful and correspond to states that are unobservable 
by that observer. For concreteness we will suppose in this Section 
that all $\Sigma_{\tau}$ are Cauchy. In Section 4 we 
demonstrate with a simple example that this approach remains useful 
in the presence of horizons.

\subsection{Particle States and S-Matrix Elements}

	Having defined $\clh^{\pm}(\tau_0)$ for any given observer, we can now define 
their {\it vacuum $| {\rm vac}_{\tau_0} \ra$ on $\Sigma_{\tau_0}$}, to be the 
Slater determinant of any basis of 
$\clh^-(\tau_0)$ (normalised so that $\la {\rm vac}_{\tau_0} | {\rm vac}_{\tau_0} 
\ra = 1$). This 
specifies $| {\rm vac}_{\tau_0} \ra$ up to an arbitrary phase factor. It is 
the state in which all negative energy degrees of freedom are full, and hence 
	is a concrete manifestation of the Dirac Sea.

	To illustrate this, let $\{ u_{i,\tau_0} ; i \in I\}$, $\{ v_{i,\tau_0} ; i \in I \}$ 
be orthonormal bases for $\clh^+(\tau_0)$ and $\clh^-(\tau_0)$ respectively, 
where $I$ is some countable index set (the 
uncountable case introduces no complications). The vacuum on $\Sigma_{\tau_0}$ can be written as:
\begin{equation} |{\rm vac}_{\tau_0} \ra = | v_{1,\tau_0} \wedge v_{2,\tau_0} 
\wedge \dots \ra \label{eq:instate} \end{equation}
and is independent of the choice of basis for $\clh^-(\tau_0)$ 
(up to a phase factor) because of the complete antisymmetry of 
the Slater determinant. The vacuum on $\Sigma_{\tau_1}$ at some `time' $\tau_1$
can similarly be written as:

\begin{equation} |{\rm vac}_{t_1} \ra = | v_{1,\tau_1} \wedge v_{2,\tau_1} 
\wedge\dots \ra \notag \end{equation} 

The observer who prepares the state `at time $\tau_0$', on 
his hypersurface of simultaneity $\Sigma_{\tau_0}$, needn't be the 
same observer who measures that state `at time $\tau_1$' (on her  
hypersurface $\Sigma_{\tau_1}$). A typical case in which these two 
observers differ is the Unruh effect, where $|\rm{vac}_{\tau_0} \ra$ 
is the Minkowski vacuum, prepared by an inertial observer, and 
$| \rm{vac}_{\tau_1} \ra$ is the Rindler vacuum, defined in the frame 
of a uniformly accelerating observer.

	The evolved state $| {\rm vac}_{\tau_0}(\tau_1) \ra$ obtained by 
evolving $|{\rm vac}_{\tau_0} \ra$ from $\Sigma_{\tau_0}$ to $\Sigma_{\tau_1}$ 
is (since $\hat{U} = \hat{U}_{1,U}$):

\begin{equation} |{\rm vac}_{\tau_0}(\tau_1) \ra = | v_{1,\tau_0}(\tau_1) 
\wedge v_{2,\tau_0}(\tau_1) \wedge \dots \ra \label{eq:forapp1} \end{equation}
 where $v_{i,\tau_0}(\tau_1)$ denotes the state (called $|\psi_{\Sigma_{\tau_0}}(\Sigma_{\tau_1})\ra$ in (\ref{eq:SSB25})) obtained from
$v_{i,\tau_0}$ by evolution to $\Sigma_{\tau_1}$. It will not in general be 
contained in $\clh^-(\tau_1)$. We will often refer to 
$|{\rm vac}_{\tau_0}(\tau_1) \ra$ as 
the `evolved vacuum', although it is not in general a vacuum state. 

	From (\ref{eq:inprod}), the vacuum-vacuum S-matrix element is simply:

\begin{equation} \la {\rm vac}_{\tau_1} | {\rm vac}_{\tau_0}(\tau_1) \ra = \det
[ \la v_{i,\tau_1} | v_{j,\tau_0}(\tau_1) \ra ] \label{Svacvac1} \end{equation}
	The probability that $|{\rm vac}_{\tau_0}(\tau_1) \ra$ will be vacuum 
at time $\tau_1$ is then $\clp_{{\rm vac} \rightarrow {\rm vac}}  = 
|\la {\rm vac}_{\tau_1} | {\rm vac}_{\tau_0}(\tau_1) \ra |^2$. Although 
this result can be derived 
by a number of methods once a particle interpretation is specified~\cite{FGS,Keif2,Sch2}, we believe 
that the present derivation (and in \cite{Me1,mythesis}) is clearer and more economical.

	States containing particles can be treated just as 
easily. For instance, a one-electron state (at time $\tau_0$) 
is of the form $u_{\tau_0} \wedge | {\rm vac}_{\tau_0} \ra$, 
and a one-positron state (at time $\tau_0$) is of the form 
$i_{v_{\tau_0}} | {\rm vac}_{\tau_0} \ra$. 
As expected, electrons are represented by the presence of positive energy 
degrees of freedom, and positrons by the absence of negative 
energy degrees of freedom (note that $v_{\tau_0} \wedge | {\rm vac}_{\tau_0} 
\ra = 0 = i_{u_{\tau_0}} | {\rm vac}_{\tau_0} \ra$ for all $u_{\tau_0} \in 
\clh^+(\tau_0)$ and $v_{\tau_0} \in \clh^-(\tau_0)$). We introduce 
the symbol $| \mbox{$\binom{i_1 i_2 \dots i_m}{j_1 j_2 
\dots j_n}$}_{\tau_0} \ra$ to denote an `in' state of $m$ particles 
(in states $u_{i_1} \dots u_{i_m}$ with $i_1 < i_2 < \dots i_m$ by 
convention), and $n$ antiparticles (corresponding to 
the absence of states $v_{j_1} \dots v_{j_n}$), prepared at time $\tau_0$. This 
state is given by:

\begin{equation} | \mbox{$\binom{i_1 i_2 \dots i_m}{j_1 j_2 \dots j_n}$}_{\tau_0} 
\ra  \equiv (-)^J | u_{i_1,\tau_0} \wedge \dots
 u_{i_m,\tau_0} \wedge v_{1,\tau_0} \wedge \dots
 \check{v}_{j_1,\tau_0} \dots \wedge \check{v}_{j_n,\tau_0} \dots \ra \label{eq:instate1} \end{equation}
	where the check over $v_{j,\tau_0}$ signifies that this degree of freedom
 is missing from the state, and $J = \frac{n}{2}(n+1) + \sum_{k=1}^n j_k$ appear as 
an unimportant sign convention. 

	The general S-matrix element can immediately be written as:
\begin{align} & \la \mbox{$\binom{i'_1 i'_2 \dots 
i'_{m'}}{j'_1 j'_2 \cdots j'_{n'}}$}_{\tau_1} | \mbox{$\binom{i_1 i_2 \dots 
i_m}{j_1 j_2 \dots j_n}$}_{\tau_0}(\tau_1) \ra \notag \\
 & = (-)^{J-J'} \det \hs{-.1} \left[ \hs{-.1} \begin{array}{cc}  
\left[ \begin{array}{ccc}
  \alpha_{i'_1 i_1} & \cdots & \alpha_{i'_1 i_m} \\ \vdots & & \vdots
  \\ \alpha_{i'_{m'} i_1}  & \cdots & \alpha_{i'_{m'} i_m} \end{array}
  \right] &  \left[ \begin{array}{ccc}  \beta_{i'_1 1} & \cdots & \binom{j_1
  \dots j_n}{missing} \\ \vdots & & \vdots \\
  \beta_{i'_{m'} 1}  & \dots & \binom{j_1 \dots j_n}{missing} \end{array} \right] \\ 
\hs{-.1} \left[ \hs{-.1} \begin{array}{ccc}  \gamma_{1 i_1}
  & \hs{-.1} \cdots \hs{-.1} & \gamma_{1 i_m} \\ 
\vdots &  &  \vdots \\
\binom{j'_1 \dots j'_{n'}}{missing} & \cdots &
  \binom{j'_1 \dots j'_{n'}}{missing}  \end{array} \hs{-.1} \right] \hs{-.2} &  
\hs{-.2} \left[ \hs{-.1} \begin{array}{ccc}
  \epsilon_{1 1} & \cdots & \hs{-.1} \binom{j_1 \dots j_n}{missing} \hs{-.1} \\
\vdots &  &  \\
  \binom{j'_1 \dots j'_{n'}}{missing} & &  \end{array} \hs{-.1} \right] \hs{-.1} 
\end{array} \hs{-.1} \right] \hs{-.1}
  \label{eq:SSBSmat}\end{align} (if $m-n = m'-n'$, and zero otherwise), where 
\bea \alpha_{i j}(\tau_1,\tau_0) = \la u_{i,\tau_1} | u_{j,\tau_0}(\tau_1) \ra & \hs{1}
\gamma_{i j}(\tau_1,\tau_0) =  \la v_{i,\tau_1} | u_{j,\tau_0}(\tau_1) \ra
\label{eq:SSB39}\\ \beta_{i j}(\tau_1,\tau_0) = \la u_{i,\tau_1} |
v_{j,\tau_0}(\tau_1) \ra & \hs{1} \epsilon_{i j}(\tau_1,\tau_0) 
= \la v_{i,\tau_1} | v_{j,\tau_0}(\tau_1) \ra \label{eq:SSB40}\eea
are the {\it time-dependent Bogoliubov coefficients}. The Bogoliubov conditions 
follow from unitarity of the `first quantized' evolution 
matrix 
$$ \bS_1(\tau_1,\tau_0) = \left[ \begin{array}{cc}  \balpha(\tau_1,\tau_0) & 
\bbeta(\tau_1,\tau_0) \\  \bgamma(\tau_1,\tau_0) & \bepsilon(\tau_1,\tau_0) \end{array} \right]$$

	The ease with which the general S-Matrix element (\ref{eq:SSBSmat}) has 
been derived contrasts with many  
conventional formulations \cite{Wa,Thal,Sch2,DeWitt}, as discussed 
in detail in \cite{Me1}. The reason is in the
concrete representation of states given 
by equations such as (\ref{eq:instate}) and (\ref{eq:instate1}), and the simple evolution equation which allows 
us to deduce equations such as (\ref{eq:forapp1}). In Canonical approaches to QFT 
in a classical background \cite{BD,Full,DeWitt} the
states are defined implicitly, by the requirement 
$a_i |\rm{vac}\ra = 0 = b_i |\rm{vac}\ra$, 
where the creation/annihilation operators $a_i$,$b_i$ are defined implicitly  
by the CAR's. The derivation of S-Matrix elements then involves   
more round about methods. One such method, analogous to that used in 
\cite{DeWitt}, is described in \cite{Me1} and contrasted with the derivation above.

\subsection{Expectation Values}

	Given an operator $\hat{A}_1(\tau):\clh \rightarrow \clh$, we can define its 
{\it physical extension} $\hat{A}_{{\rm phys}}(\tau): \clf_{\wedge}(\clh) \rightarrow 
\clf_{\wedge}(\clh)$ by:
\begin{equation} \hat{A}_{{\rm phys}}(\tau) = \hat{A}_H(\tau) - \la {\rm vac}_{\tau} | \hat{A}_H(\tau) | {\rm vac}_{\tau} \ra \hat{1} \label{eq:vacsub1}  \end{equation}

	This is the relativistic equivalent of the `one-particle operator' 
of multiparticle quantum mechanics \cite{MYS}, and is expressible as a 
normal-ordered bilinear of the field operator $\hat{\psi}(x)$. This 
vacuum subtraction is equivalent to normal 
ordering with respect to the particle 
interpretation {\it at the time of measurement}. This choice is also made in  
previous `Hamiltonian diagonalisation' procedures \cite{GM1,GM2,MMS,GMM2,GMM}, and uniquely  
guarantees the positive definiteness of $\hat{H}_{{\rm phys}}(\tau)$ while maintaining 
$\la {\rm vac}_{\tau} | \hat{H}_{{\rm phys}}(\tau) | {\rm vac}_{\tau} \ra = 0$. This 
is discussed in more detail in \cite{Me1}, where the relation of our 
approach to Hamiltonian Diagonalisation is also described.

	The expectation value of $\hat{A}_{{\rm phys}}(\tau)$ in the physical 
vacuum $| {\rm vac}_{\tau} \ra$ at time $\tau$ is zero by construction.  
Its expectation value in the `evolved vacuum' $| {\rm vac}_{\tau_0}(\tau) \ra$ 
is in general non-zero, and takes the form 
\begin{align} \la {\rm vac}_{\tau_0}(\tau_1) & | \hat{A}_{{\rm phys}}(\tau_1) 
| {\rm vac}_{\tau_0}(\tau_1) \ra \label{eq:vac1} \\
 & \hs{-.5} = \sum_{i=1}^{N} \la v_{i,\tau_0}(\tau_1) | \hat{A}_1(\tau_1) 
| v_{i,\tau_0}(\tau_1) \ra - \sum_{i=1}^{N} 
\la v_{i,\tau_1} | \hat{A}_1(\tau_1) | v_{i,\tau_1} \ra  \label{eq:vacsub} \\
	& \hs{-.5} = {\rm Trace}(\bbeta \bbeta^{\dagger} \bA^{++} - \bgamma
	\bgamma^{\dagger} \bA^{--} + \bepsilon \bbeta^{\dagger}
	\bA^{+-} + \bbeta \bepsilon^{\dagger} \bA^{-+})
	\label{eq:SSB40.1} \end{align} 
where we have defined:
\begin{align} \bA^{++}_{j k} \equiv \la u_{j,\tau_1} | \hat{A}_1(\tau_1) | u_{k,\tau_1} \ra 
 \hs{1} & \bA^{--}_{j k} \equiv \la v_{j,\tau_1} | \hat{A}_1(\tau_1) | v_{k,\tau_1} \ra \notag \\ 
\bA^{+-}_{j k} \equiv \la u_{j,\tau_1} | \hat{A}_1(\tau_1) |
  v_{k,\tau_1} \ra \hs{.2} \mbox{ and } & \bA^{-+}_{j k} \equiv \la
  v_{j,\tau_1} | \hat{A}_1(\tau_1) | u_{k,\tau_1} \ra \label{eq:SSB40.2} \\
 & \hs{.8} = \overline{\bA^{+-}_{k j}} \mbox{ if } \hat{A}_1 
\mbox{ is Hermitian } \notag \end{align}

The relation $(\bepsilon \bepsilon^{\dagger})_{k j} - \delta_{k j} =
  - (\bgamma \bgamma^{\dagger})_{k j}$ has been used in deriving 
(\ref{eq:SSB40.1}) from (\ref{eq:vacsub}). This step 
  relies on the fact that we are vacuum
  subtracting with respect to the vacuum {\it at the time of 
measurement}. Notice that if $\hat{A}_1$ is conserved at the level of the Dirac
equation, then:
$$ \la {\rm vac}_{\tau_0}(\tau_1) | \hat{A}_{{\rm phys}} | {\rm vac}_{\tau_0}(\tau_1) \ra = 
{\rm Trace}(\bA^{--}(\tau_0) - \bA^{--}(\tau_1)) $$
	which can be non-zero even when $\hat{A}_{1}$ is independent of
	time, because of the varying particle interpretation. Herein lies 
an elegant physical description of quantum anomalies, as outlined  
in \cite{Me1}. A treatment of the axial anomaly, described in terms of this
physical mechanism, has been given in \cite{AxAn} and in \cite{Jackiw}.

	The derivation of $\la F_{\tau_0}(\tau_1) |
  \hat{A}_{{\rm phys}}(\tau_1) | F_{\tau_0}(\tau_1) \ra$ for an arbitrary state $| F_{\tau_0}(\tau_1) \ra$
  is identical to the derivation of (\ref{eq:SSB40.1}), with result:

\begin{align} & \la \mbox{$\binom{i_1 i_2 \dots i_m}{j_1 j_2 \dots j_n}$}_{\tau_0}(\tau_1)|
 \hat{A}_{{\rm phys}}(\tau_1) | \mbox{$\binom{i_1 i_2 \dots i_m}{j_1 j_2 
\dots j_n}$}_{\tau_0}(\tau_1) \ra \notag \\
 & \hs{1} = \sum_{k=1}^{m} \la u_{i_k,\tau_0}(\tau_1) | \hat{A}_1(\tau_1) |
 u_{i_k,\tau_0}(\tau_1) \ra - \sum_{k=1}^{n} \la v_{j_k,\tau_0}(\tau_1) | \hat{A}_1(\tau_1)
 | v_{j_k,\tau_0}(\tau_1) \ra \notag \\
 & \hs{2} + \la {\rm vac}_{\tau_0}(\tau_1) | \hat{A}_{{\rm phys}}(\tau_1)
 | {\rm vac}_{\tau_0}(\tau_1) \ra \label{eq:SSB40.12} \end{align}

We now consider some simple examples of physical extension. \newline

	The charge operator is easily recognised as the physical extension of 
the unit operator on $\clh$, $\hat{Q} = e \hat{1}_{{\rm phys}}$. This is 
 appropriate, since the norm $\la \psi | \hat{1}_1 | \psi \ra$ of a
 state $\psi \in \clh$ is the Noether charge of the Dirac Lagrangian that is 
conjugate to changes in phase.  Charge conservation 
now follows directly from the fact that evolution is
 grade-preserving.

	The operator $\hat{N}_{{\rm phys}}(\tau)$ that represents the number of 
particles (including antiparticles) present at time $\tau$ is the physical 
extension of $\hat{N}_1(\tau) = \hat{P}^+_{\tau} - \hat{P}^-_{\tau}$, where 
$\hat{P}^{\pm}_{\tau}:\clh \rightarrow \clh^{\pm}(\tau)$ are the projection operators 
onto $\clh^{\pm}(\tau)$. Clearly $\hat{N}_1(\tau_0)$ commutes with $\hat{H}_1(\tau_0)$, but does not
in general commute with time evolution (since it does not commute with 
$\hat{H}_1(\tau)$ for $\tau \neq \tau_0$). $\hat{N}_{{\rm phys}}(\tau_0)$ inherits both of 
these properties. Therefore the number operator $\hat{N}_{{\rm phys}}(\tau_0)$
represents a well-defined physical observable, but one which is not
conserved. When acting on states in standard form, it gives:

\begin{equation} \hat{N}_{{\rm phys}}(\tau_0) | \mbox{$\binom{i_1 i_2 \dots i_m}{j_1 j_2 
\dots j_n}$}_{\tau_0} \ra = (m + n) | \mbox{$\binom{i_1 i_2 \dots i_m}{j_1 j_2 
\dots j_n}$}_{\tau_0} \ra 
 \label{eq:SSB22}\end{equation}
	so that it is positive definite and has integer eigenvalues. 
From (\ref{eq:SSB40.1}), the expectation value of $\hat{N}_{{\rm phys}}(\tau_1)$ in the `evolved vacuum' 
$| {\rm vac}_{\tau_0}(\tau_1) \ra$ is given by:
\begin{align} N_{{\rm vac},\tau_0}(\tau_1) & \equiv \la {\rm vac}_{\tau_0}(\tau_1)| 
\hat{N}_{{\rm phys}}(\tau_1) | {\rm vac}_{\tau_0}(\tau_1) \ra \notag \\
 & = {\rm Trace}( \bbeta \bbeta^{\dagger} + \bgamma \bgamma^{\dagger}) \label{eq:ntot1} \\
 & = \sum_i \{ N^+_{i,\tau_0}(\tau_1) + N^-_{i,\tau_0}(\tau_1) \} \notag \end{align}
where $N^+_{i,\tau_0}(\tau_1) = (\bbeta \bbeta^{\dagger})_{i i} = \sum_j | 
\la u_{i,\tau_1}| v_{j,\tau_0}(\tau_1) \ra|^2  $ is the expectation 
value of the physical extension of $| u_{i,\tau_1} \ra \la u_{i,\tau_1}|$, and 
represents 
the probability that the degree of freedom $u_{i,\tau_1}$ is occupied in 
$| {\rm vac}_{\tau_0}(\tau_1) \ra$, i.e. that particle $i$ is present. 
$N^-_{i,\tau_0}(\tau_1) = (\bgamma \bgamma^{\dagger})_{i i} = \sum_j | 
\la v_{i,\tau_1}| u_{j,\tau_0}(\tau_1) \ra|^2$ is the expectation 
value of the physical extension of $- | v_{i,\tau_1} \ra \la v_{i,\tau_1}|$;
it represents the probability that the degree of freedom $v_{i,\tau_1}$ is 
unoccupied in $| {\rm vac}_{\tau_0}(\tau_1) \ra$,  i.e. that antiparticle $i$ is present. 
The Bogoliubov conditions imply ${\rm Trace}( \bbeta \bbeta^{\dagger}) = 
{\rm Trace}( \bgamma \bgamma^{\dagger})$, which again expresses charge conservation.
	By defining the projection operators $\hat{P}^{\pm}_{\tau_0}(\tau)$ by:
\begin{align} \hat{P}^{+}_{\tau_0}(\tau) & \equiv \sum_i |u_{i,\tau_0}(\tau) \ra \la u_{i,\tau_0}(\tau)| 
= \hat{U}_1(\tau,\tau_0) \hat{P}^+_{\tau_0} \hat{U}^{\dagger}_1(\tau,\tau_0) 
 \label{eq:p+} \\
\hat{P}^{-}_{\tau_0}(\tau) & \equiv \sum_i |v_{i,\tau_0}(\tau) \ra \la v_{i,\tau_0}(\tau)| = 
\hat{U}_1(\tau,\tau_0) \hat{P}^-_{\tau_0} \hat{U}^{\dagger}_1(\tau,\tau_0) \label{eq:p-}\end{align}
	we can write $N^{\pm}_{i,\tau_0}(\tau_1)$ as:
\begin{equation} N^{+}_{i,\tau_0}(\tau_1) = \la u_{i,\tau_1} | \hat{P}^{-}_{\tau_0}(\tau_1) 
| u_{i,\tau_1} \ra \hs{.5} 
N^{-}_{i,\tau_0}(\tau_1) = \la v_{i,\tau_1} | \hat{P}^{+}_{\tau_0}(\tau_1) 
| v_{i,\tau_1} \ra \label{eq:nproj} \end{equation}

	We can rewrite (\ref{eq:SSB40.1}) in terms of $\hat{P}^{\pm}_{\tau_0}(\tau)$ as:

\begin{align}
\la \rm{vac}_{\tau_0}(\tau_1) | \hat{A}_{\rm{phys}}(\tau_1) | \rm{vac}_{\tau_0}(\tau_1) \ra
& = \rm{Trace}( \hat{A}_1(\tau_1) (\hat{P}^{-}_{\tau_0}(\tau_1) - \hat{P}^{-}_{\tau_1})) \label{eq:fluc11}\\
& = - \rm{Trace}( \hat{A}_1(\tau_1) (\hat{P}^{+}_{\tau_0}(\tau_1) - \hat{P}^{+}_{\tau_1}))
\notag \end{align}
where we have used $\hat{P}^{+}_{\tau_0}(\tau_1) + \hat{P}^{-}_{\tau_0}(\tau_1) 
= \hat{1} = \hat{P}^{+}_{\tau_1} + \hat{P}^{-}_{\tau_1}$. This result can be 
generalised to any state $|F\ra$ of the form $|F(\tau)\ra = |\psi_1(\tau) \wedge 
\psi_2(\tau) \dots \wedge \psi_n(\tau) \ra$ where $\psi_1(\tau),\psi_2(\tau),\dots \psi_n(\tau)$
are orthonormal (which includes all $| \mbox{$\binom{i_1 i_2 \dots i_m}{j_1 j_2 
\dots j_n}$}_{\tau_0}(\tau) \ra$). In this case, define:
\begin{align} \hat{P}^{-}_{|F\ra}(\tau) & \equiv \sum_{i \in I} |\psi_{i}(\tau) \ra \la \psi_{i}(\tau)| \notag \\
\mbox{ and } \hat{P}^{+}_{|F\ra}(\tau) & = \equiv \sum_{i \notin I} |\psi_{i}(\tau) \ra \la \psi_{i}(\tau)| = \hat{1} - \hat{P}^{-}_{|F\ra}(\tau) \notag \end{align}
	where $\sum_{i \in I}$ runs over $\{ \psi_1, \dots \psi_n \}$, and $\sum_{i \notin I}$ runs over the orthogonal complement of this. Then (\ref{eq:SSB40.12}) can be written as:

\begin{align}
\la F(\tau) | 
\hat{A}_{\rm{phys}}(\tau) | F(\tau) \ra & = \rm{Trace}( \hat{A}_1(\tau) 
(\hat{P}^{-}_{|F\ra}(\tau) - \hat{P}^{-}_{\tau})) \notag \\
& = - \rm{Trace}( \hat{A}_1(\tau) (\hat{P}^{+}_{|F\ra}(\tau) - \hat{P}^{+}_{\tau})) \label{eq:fluc12} \\
& = \rm{Trace}( \hat{A}_1(\tau) (\hat{P}^{-}_{|F\ra}(\tau) \hat{P}^{+}_{\tau}
- \hat{P}^{+}_{|F\ra}(\tau) \hat{P}^{-}_{\tau})) \notag 
\end{align}
	The relation of these projection operators to the various Greens functions of the theory, and to the first order `Dirac density matrix' of multiparticle quantum mechanics (see for instance \cite{MYS}, pp 9-10) is described in the Appendix.

\subsection{Fluctuations}

Fluctuations in expectation values can be calculated using (\ref{eq:prop3c}). In the notation of the previous subsection, we can write (\ref{eq:prop3c} as:

\begin{align} \la F(\tau) & |(\hat{A}_{\rm{phys}})^2 | F(\tau) \ra  
- (\la F(\tau) | \hat{A}_{\rm{phys}} | F(\tau) \ra)^2 = 
\sum_{i \in I, j \notin I}  | \la \psi_i(\tau) | \hat{A}_1 | \psi_j(\tau) \ra |^2 \notag \\
 & = {\rm Trace}(\hat{P}^-_{|F\ra}(\tau) \hat{A}_1(\tau) \hat{P}^+_{|F\ra}(\tau) \hat{A}_1(\tau)) = - \half \rm{Trace}([\hat{P}^-_{|F\ra}(\tau),\hat{A}_1(\tau)]^2)  \notag \end{align} 
	In the evolved vacuum $| \rm{vac}_{\tau_0}(\tau) \ra$ this becomes:

\begin{align} \la \hat{A}^2 \ra & \equiv \la {\rm vac}_{\tau_0}(\tau) | 
(\hat{A}_{{\rm phys}})^2 | {\rm vac}_{\tau_0}(\tau) \ra
 - \la {\rm vac}_{\tau_0}(\tau) | \hat{A}_{{\rm phys}} 
| {\rm vac}_{\tau}(\tau) \ra^2 \notag \\ 
& = \sum_{i,j} | \la v_{i,\tau_0}(\tau) | \hat{A}_1(\tau) 
|u_{j,\tau_0}(\tau) \ra|^2 \notag \\
 & = {\rm Trace}(\hat{P}^-_{\tau_0}(\tau) \hat{A}_1(\tau) \hat{P}^+_{\tau_0}(\tau) \hat{A}_1(\tau)) \notag \\
&  = \rm{Trace}(\bA_F \bA_F^{\dagger}) \label{eq:fluc10} \\
\mbox{ where } \bA_F & \equiv \bbeta^{\dagger} \bA^{+ +} \balpha + \bepsilon^{\dagger} \bA^{- -} \bgamma + \bbeta^{\dagger} \bA^{+ -} \bgamma + \bepsilon^{\dagger} \bA^{- +} \balpha \label{eq:fluc20} \end{align}

	Consider for example fluctuations in $\hat{N}^+(\tau)$, the 
total number of particles (not including antiparticles) in the evolved 
vacuum. Now $\hat{N}_1^+(\tau) = \hat{P}^{+}_{\tau}$ so that $\bA_F = \bbeta^{\dagger} \balpha$, and:
\begin{equation} \la (\hat{N}^{+})^2 \ra = \rm{Trace}(\bbeta 
\bbeta^{\dagger} (1 - \bbeta \bbeta^{\dagger})) \label{eq:fluc28} \end{equation}
	Consider a spatially uniform case, with $\beta_{\bp \lambda; \bq \sigma}(\tau,\tau_0) = \beta_{\bp} \delta_{\lambda \sigma} (2 \pi)^3 \delta(\bp - \bq)$. Then:
\begin{align} \frac{N^+_{\rm{vac},\tau_0}(\tau)}{V} & = 2 
\int \frac{d^3 \bp}{(2 \pi)^3} |\beta_{\bp}|^2 \label{eq:fluc30} \\ 
\mbox{while} \hs{2} \la \left(\frac{N^+_{\rm{vac},\tau_0}(\tau)}{V}\right)^2 \ra & 
= \frac{2}{V} \int \frac{d^3 \bp}{(2 \pi)^3} |\beta_{\bp}|^2 (1 - 
|\beta_{\bp}|^2) \label{eq:fluc40} \end{align} 

	This result suggests that for large volumes, and for 
$|\beta_{\bp}|^2 << 1$, the fluctuations in the average particle density 
in a volume $V$ vary inversely with $V$, and that in a given volume the 
rms error in the total number of particles is proportional to the square 
root of the number of particles, as would be expected for a weakly 
interacting system. The same result holds for the total number of 
antiparticles, or for the total number of pairs. (The fluctuations in 
the total charge are zero, due to the Bogoliubov conditions.) However, 
`$V$' in (\ref{eq:fluc30}) and (\ref{eq:fluc40})
is formally infinite; $V = (2 \pi)^3 \delta^3(\b0)$. Before we can confirm 
our interpretation of (\ref{eq:fluc30}) and (\ref{eq:fluc40}) we 
must consider finite volume measurements.

\subsection{Finite Volume Measurements}

	To measure the quantity $A$ in a volume $V$ (on $\Sigma_{\tau}$) it 
seems reasonable to consider the operator $\hat{A}_{V,\rm{phys}}$ obtained 
by Hermitian extension and vacuum subtraction from the operator: 

\begin{equation} \hat{A}_{V,1} \equiv \half (\hat{\theta}_V \hat{A}_1 + \hat{A}_1 
\hat{\theta}_V ) \mbox{where } \hat{\theta}_V \psi(x_{\tau}) \equiv \begin{cases}  
 \psi(x_{\tau}) & x_{\tau} \in V \\
 0 & x_{\tau} \notin V \\ \label{eq:fluc42} \end{cases} \end{equation} 

	where $x_{\tau}$ is shorthand for $x|_{\Sigma_{\tau}}$ (the 
restriction of $x$ onto $\Sigma_{\tau}$). The matrix elements of 
$\hat{A}_{V,1}$ are related to those of $\hat{A}_{1}$ by restricting the 
volume integral to $V$, taking the appropriate combination of surface 
terms to ensure Hermiticity. In the Canonical approach, $\hat{A}_{V,H}$
is similarly obtained by restricting the bilinear $\int_{\Sigma_{\tau}} e(x) 
\half [ \overline{\hat{A}_1 \hat{\psi}(x)} \gamma^{\mu}(x) \hat{\psi}(x) + 
\overline{\hat{\psi}(x)} \gamma^{\mu}(x) \hat{A}_1 \hat{\psi}(x) ] d \Sigma_{\mu}$ 
to the volume $V$ (here $\hat{\psi}(x)$ is the field operator). 

	As defined, however, $\hat{A}_{V,\rm{phys}}$ gives rise to some serious problems:
\begin{enumerate}

\item The fluctuations in $\hat{N}^+_{V,\rm{phys}}(\tau)$, 
$\hat{N}^-_{V,\rm{phys}}(\tau)$, $\hat{Q}_{V,\rm{phys}}$ and 
$\hat{H}_{V,\rm{phys}}(\tau)$  are infinite for any 
finite $V$, even in the physical vacuum $|\rm{vac}_{\tau} \ra$, and even 
for an inertial observer in flat empty space!  To see this, note that:
\begin{align} \la {\rm vac}_{\tau} | 
(\hat{N}^{\pm}_{V,{\rm phys}})^2 | {\rm vac}_{\tau} \ra
 & = \sum_{i j} | \la u_{j,\tau} | v_{i, \tau} \ra_V |^2 \label{eq:Fluc45} \\
\la {\rm vac}_{\tau} | 
(\hat{H}_{V,{\rm phys}})^2 | {\rm vac}_{\tau} \ra
 & = \sum_{i j} (E^+_j - E^-_i)^2 | \la u_{j,\tau} | v_{i, \tau} \ra_V |^2
\label{eq:Fluc50} \end{align}
where $\la u_{j,\tau} | v_{i, \tau} \ra_V$ signifies that the integral in the 
inner product has been restricted to $V$ (it would be zero otherwise). For 
convenience $| u_{j,\tau} \ra$ is chosen to be an 
eigenstate with eigenvalue $E^+_j$ and $| v_{i,\tau} \ra$ is an 
eigenstate with eigenvalue $- E^-_i$. For an inertial observer in flat 1+1 
dimensional space, and $V$ being the region $|x| < \frac{L}{2}$, this becomes:
\begin{align} \la {\rm vac}_{\tau} | (\hat{H}_{V,{\rm phys}})^2 & | 
{\rm vac}_{\tau} \ra  = \int \frac{d p}{2 \pi} \frac{d q}{2 \pi} F(p,q) \notag \\
\hs{-.5} \mbox{ where } F(p,q) \equiv (E_p - E_q)^2 & \frac{(E_p + p)(E_q + q)}{2 E_p 
2 E_q} (1 - \frac{m^2}{(E_p + p)(E_q + q)})^2 \frac{ \sin^2[(p+q)L]}{(p+q)^2} 
\notag \end{align}
	which is easily seen to diverge (the divergence is just as bad in 3+1 dimensions).

\item The expectation values of $\hat{N}^{\pm}_{V,\rm{phys}}(\tau)$ in the 
evolved vacuum $|\rm{vac}_{\tau_0}(\tau) \ra$ are given by:
\begin{align} \la \rm{vac}_{\tau_0}(\tau) |\hat{N}^{+}_{V,\rm{phys}}(\tau)| 
\rm{vac}_{\tau_0}(\tau) \ra & = \rm{Trace} (\bbeta \bbeta^{\dagger} 
\la u_{i,\tau} | u_{j,\tau} \ra_V + \Re(\bepsilon \bbeta^{\dagger} \la u_{i,\tau} 
| v_{j,\tau} \ra_V)) \label{eq:fluc60} \\
\la \rm{vac}_{\tau_0}(\tau) |\hat{N}^{-}_{V,\rm{phys}}(\tau)| 
\rm{vac}_{\tau_0}(\tau) \ra & = \rm{Trace} (\bgamma \bgamma^{\dagger} 
\la v_{i,\tau} | v_{j,\tau} \ra_V - \Re(\bepsilon \bbeta^{\dagger} \la u_{i,\tau} 
| v_{j,\tau} \ra_V)) \label{eq:fluc61} \end{align}
	The first terms represent the obvious contribution from the sum 
over created particles of the probability distribution of each created particle. 
The final term has no obvious interpretation. It is comparable in 
magnitude to the first term, but is not positive 
definite, and it affects $\la \rm{vac}_{\tau_0}(\tau) 
|\hat{N}^{+}_{V,\rm{phys}}(\tau)| \rm{vac}_{\tau_0}(\tau) \ra$ with opposite 
sign to $\la \rm{vac}_{\tau_0}(\tau) |\hat{N}^{-}_{V,\rm{phys}}(\tau)| 
\rm{vac}_{\tau_0}(\tau) \ra$.

\end{enumerate}

	All these problems stem from the fact that $\hat{\theta}_V$ does 
not commute with $\hat{P}^{\pm}_{\tau}$. This is the well-known 
fact that particle states cannot be confined to a finite volume 
without introducing a negative-energy component. Consequently, even if 
$[ \hat{A}_1,\hat{P}^{\pm}_{\tau}] = 0$ the same will not be true of 
$\hat{A}_{V,1}$. Hence, even if $|\rm{vac}_{\tau}\ra$ is an eigenstate of 
$\hat{A}_{\rm{phys}}$, it will not in general be an eigenstate of 
$\hat{A}_{V,\rm{phys}}$ for finite $V$. This can be seen in the Canonical 
approach by considering the transition from operators expressed in 
terms of $\hat{\psi}(x)$ to operators expressed in terms of 
creation/anihilation operators. The absence of terms proportional to 
$a^{\dagger}_i b^{\dagger}_j$ relies on the relation $\la u_i | \hat{A}_1 | v_j 
\ra = 0$. However, when the integral is restricted to a finite 
volume this matrix element can be non-zero even if $|v_j\ra$ is an eigenstate of $\hat{A}_1$. It is these finite volume overlaps between positive and negative 
energy states that contribute to (\ref{eq:Fluc45}) and (\ref{eq:Fluc50}), 
and lead to the unphysical second term in (\ref{eq:fluc60}) and 
(\ref{eq:fluc61}).

	This problem is easily overcome. Define:
\begin{align}
\hat{\theta}^{\tau}_V & \equiv \hat{P}^{+}_{\tau} \hat{\theta}_V  
\hat{P}^{+}_{\tau} +   \hat{P}^{-}_{\tau} \hat{\theta}_V   \hat{P}^{-}_{\tau}
\label{eq:fluc70} \\
\mbox{and } \hat{A}_{V,1} & \equiv \half (\hat{\theta}^{\tau}_V \hat{A}_1 + \hat{A}_1 \hat{\theta}^{\tau}_V) \label{eq:fluc71} \end{align}
	The expressions considered in (\ref{eq:fluc42}) - (\ref{eq:fluc61}) 
will henceforth be denoted $\hat{A}^{\rm{naive}}_{V,1}$, 
$\hat{A}^{\rm{naive}}_{V,\rm{phys}}$, etc. $\hat{A}_{V,1}$ now has 
the property that $[ \hat{A}_{V,1},\hat{P}^{\pm}_{\tau}] = 0$ (for all $V$) 
if and only if $[ \hat{A}_1,\hat{P}^{\pm}_{\tau}] = 0$. Hence, if 
$|\rm{vac}_{\tau} \ra$ is an eigenstate of $\hat{A}_{\rm{phys}}$, then it 
is an eigenstate of $\hat{A}_{V,\rm{phys}}$ for all $V$, so that the 
fluctuations of $A$ in any finite volume in the physical vacuum 
$|\rm{vac}_{\tau} \ra$ will be zero. This applies for instance to 
$\hat{N}^{\pm}(\tau)$, $\hat{Q}$ and $\hat{H}$, and resolves problem 1 above. 

	If $[\hat{A}_1,\hat{P}^{\pm}_{\tau}] = 0$, then:
\begin{equation} 
\la {\rm vac}_{\tau_0}(\tau_1) | \hat{A}_{V, {\rm phys}}(\tau_1) 
| {\rm vac}_{\tau_0}(\tau_1) \ra = {\rm Trace}(\bbeta \bbeta^{\dagger} 
\bA^{++}_V - \bgamma \bgamma^{\dagger} \bA^{--}_V) \notag \end{equation}

where $\bA^{++}_{V,j k} \equiv \la u_{j,\tau_1} | 
\hat{A}^{\rm{naive}}_{1,V}(\tau_1) | u_{k,\tau_1} \ra$ and 
$\bA^{--}_{j k} \equiv \la v_{j,\tau_1} | \hat{A}^{\rm{naive}}_{1,V}(\tau_1) 
| v_{k,\tau_1} \ra$. The terms describing finite-volume overlaps between 
positive and negative energy states have been removed, resolving  
problem 2. For instance, we can now write:

\begin{align} N^-_{i,V,\tau_0}(\tau) & = \Re(\la v_{i,\tau} | \hat{\theta}_V 
\hat{P}^-_{\tau} \hat{P}^+_{\tau_0}(\tau) |v_{i,\tau} \ra = \sum_k 
\Re[\rm{Trace}((\bgamma \bgamma^{\dagger})_{i k} \la v_{k,\tau} 
| v_{i,\tau} \ra_V)] \notag \\
N^+_{i,V,\tau_0}(\tau) & = \Re(\la u_{i,\tau} | \hat{\theta}_V 
\hat{P}^+_{\tau} \hat{P}^-_{\tau_0}(\tau) |u_{i,\tau} \ra = \sum_k 
\Re[\rm{Trace}((\bbeta \bbeta^{\dagger})_{i k} \la u_{k,\tau} 
| u_{i,\tau} \ra_V)] \notag \end{align}
	from which it follows that:
\begin{align} N^-_{V,\tau_0}(\tau) & = \rm{Trace} (\bgamma \bgamma^{\dagger} \la v_{k,\tau} | v_{i,\tau} \ra_V) \notag \\
& = \int_V e(x) J^{- \mu}_{\tau_0}(x) d \Sigma_{\mu} \hs{.5} 
= \int_V n^-_{\tau_0}(\bx) d^3 \bx      \label{eq:numdens-} \\
\mbox{ where } \hs{1} J^{- \mu}_{\tau_0}(x) & \equiv \sum_{i k} (\bgamma \bgamma^{\dagger})_{i k} \bar{v}_{k,\tau(x)}(x) \gamma^{\mu} v_{i,\tau(x)}(x) \notag \\
\mbox{ and } \hs{1} n^-_{\tau_0}(\bx) & \equiv e(x_{\tau}) J^{- \mu}_{\tau_0}(x_{\tau})
\frac{\partial_{\mu} \tau}{\partial_0 \tau}|_{x = x_{\tau}} \mbox{ where } 
x^{\mu}_{\tau} = (t_{\tau}(\bx), \bx) \notag \\
N^+_{V,\tau_0}(\tau) & = {\rm Trace} (\bbeta \bbeta^{\dagger} \la u_{k,\tau} | u_{i,\tau} \ra_V) \notag \\
& = \int_V e(x) J^{+ \mu}_{\tau_0}(x) d \Sigma_{\mu} \hs{.5} 
= \int_V n^+_{\tau_0}(x) d^3 \bx \label{eq:numdens+} \\
\mbox{ where } \hs{1} J^{+ \mu}_{\tau_0}(x) & \equiv \sum_{i k} (\bbeta \bbeta^{\dagger})_{i k} \bar{u}_{k,\tau(x)}(x) \gamma^{\mu} u_{i,\tau(x)}(x) \notag \\
\mbox{ and } \hs{1} n^+_{\tau_0}(\bx) & \equiv e(x_{\tau}) J^{+ \mu}_{\tau_0}(x_{\tau})
\frac{\partial_{\mu} \tau}{\partial_0 \tau}|_{x = x_{\tau}} \notag \end{align}
	Then $N_{V,\tau_0}(\tau) = \int_V e(x) J^{+,\mu}_{\tau_0}(x) + J^{-,\mu}_{\tau_0}(x) d \Sigma_{\mu}$ and $Q_{V,\tau_0}(\tau) = e \int_V e(x) J^{+,\mu}_{\tau_0}(x) - J^{-,\mu}_{\tau_0}(x) d \Sigma_{\mu}$.
	The fluctuations in $N^{+}_{V,\tau_0}(\tau)$ are given by:
\begin{equation} \la (\hat{N}^+_{V})^2 \ra = \rm{Trace}((\bbeta 
\bbeta^{\dagger})_{i j} \la u_{j,\tau} | u_{k,\tau} \ra_V (1 - \bbeta 
\bbeta^{\dagger})_{k l} \la u_{l,\tau} | u_{m,\tau} \ra_V) \label{eq:fluc80} 
\end{equation}
	which clearly reduces to (\ref{eq:fluc28}) when $V$ covers all 
of $\Sigma_{\tau}$. In the spatially uniform case, with 
$\beta_{\bp \lambda; \bq \sigma}(\tau,\tau_0) = \beta_{\bp} 
\delta_{\lambda \sigma} (2 \pi)^3 \delta(\bp - \bq)$, it is straightforward to 
show that (\ref{eq:fluc40}) also holds approximately for large finite $V$, 
confirming the interpretation of (\ref{eq:fluc40}) given at the 
end of Section 3.4. Similarly, the fluctuations in 
$N^{-}_{V,\tau_0}(\tau)$ are given by:
\begin{equation} \la (\hat{N}^-_{V})^2 \ra = \rm{Trace}((\bgamma 
\bgamma^{\dagger})_{i j} \la v_{j,\tau} | v_{k,\tau} \ra_V (1 - \bgamma 
\bgamma^{\dagger})_{k l} \la v_{l,\tau} | v_{m,\tau} \ra_V) 
\label{eq:fluc90} \end{equation}
	In the absence of an electromagnetic field,  
Charge Conjugation invariance implies that $N^{+}_{V,\tau_0}(\tau) = N
^{-}_{V,\tau_0}(\tau)$ and $\la (\hat{N}^+_{V})^2 \ra = \la (\hat{N}^+_{V})^2 
\ra$ for all $V$. Discrete symmetries will be addressed in a future publication.

	We have described, both here and in previous publications 
\cite{mythesis,Me2}, the utility of a finite-time particle interpretation. 
In \cite{Me3} we demonstrated with examples how it becomes 
possible to track the particle production process, finding not only how many 
particles are created, but also when they are created, and how they 
behave after their creation. Having now defined finite-volume operators 
which have well-defined fluctuations, we can also say (with controllable 
precision) where the particles are created. In a future  
publication we will show that this makes it possible to treat `particle creation' and `vacuum 
polarisation' effects within the same framework (\cite{GMR} presents a similar
 treatment for electrostatic fields). In particular, for an electromagnetic 
potential step, modes with evanescent contributions give rise to a 
nonzero charge distribution in the vicinity of the barrier (on a length scale 
$\lambda_c = \frac{\hbar}{m c}$), providing a vacuum polarisation which 
partially screens the barrier. Meanwhile, the `Klein modes', present only 
for a potential step larger than $2 m c^2$, correspond to the standard `particle 
creation' effect, with created plane wave states (persisting to spatial 
infinity). Conventional `tunneling' methods, based on calculating 
transmission/reflection coefficients, describe only those modes which 
persist as plane waves either side of the barrier. Methods based only 
on $\hat{N} = \sum_i \{ a^{\dagger}_i a_i + b^{\dagger}_i b_i \}$, pick 
up the total number of created particles, but this generally must 
be divided by the infinite spatial volume $(2 \pi)^3 \delta(\b0)$ to 
obtain an average particle density. Again, only those modes contribute which 
persist to spatial infinity.

	Further details of our approach to fermionic QFT are 
given in \cite{Me1,mythesis}. We now study a simple example 
involving a particle horizon; the well-known Unruh effect, concerning a 
uniformly accelerating observer in flat Minkowski spacetime.

\section{Rindler Space: A Simple Horizon}

The Unruh effect \cite{Unruh,Davies} is one of the most 
studied and most cited examples of particle 
creation, and demonstrates more than any other the fact that 
the concept of particle is observer-dependent. Useful references 
are \cite{BD,Full,GMR} and the further references therein. An early treatment 
of fermions in Rindler space is \cite{SMG} (presented again in the same authors  
text \cite{GMR}), while \cite{Tak} presents a very thorough review 
of fermions in Rindler space. We now consider this problem using the present 
formalism. We begin by showing that the radar time of
 a uniformly accelerating observer is indeed Rindler time, and that 
the eigenstates of $\hat{H}_1$ are stationary states of Rindler 
time. This demonstrates the consistency of this approach with 
standard derivations. We complete the derivation in 1+1 dimensions for  
both massive and massless particles. Although the rest of this 
derivation is quite standard, we have also described the spatial 
distribution of Rindler particles, which has received 
little attention to date.

\subsection{Radar Time for Accelerating Observers}

	Consider first the 1+1 dimensional case. We have 
\begin{equation} x^{\mu}_{(\gamma)}(\tau) \equiv (t(\tau) , z(\tau)) = 
(\frac{\sinh(a \tau)}{a} , \frac{\cosh(a \tau)}{a}) \label{eq:5.1.1} \end{equation}

\begin{figure}[h]
\center{\epsfig{figure=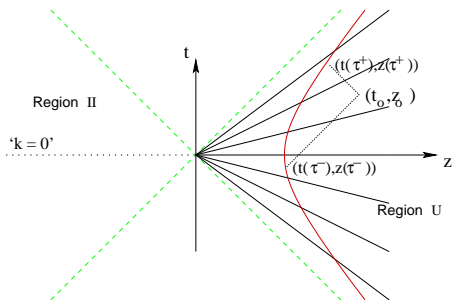, width=6cm}}
\caption{{\footnotesize Hypersurfaces of simultaneity of a uniformly accelerating observer.}}
\end{figure}

	Consider a point $(t_o,z_o)$ to the right of the observer's worldline, as shown 
in figure 2. Clearly, the $\tau^{\pm}$ for this point must satisfy:
\begin{equation} t(\tau^+) - t_o = z_o - z(\tau^+) \hs{1} t_o - t(\tau^-) = 
z_o - z(\tau^-) \label{eq:deriv1} \end{equation}
From (\ref{eq:5.1.1}),
\begin{equation} e^{a \tau^+} = a(z_o + t_o) \hs{1} e^{- a \tau^-} = a(z_o - t_o)
\notag \end{equation}
	from which we deduce that $\tau = \frac{1}{2a} 
\log(\frac{z_o + t_o}{z_o - t_o})$. For points $(t_o,z_o)$ to the 
left of the observers worldline (but still in Region U), the roles 
of $\tau^+$ and $\tau^-$ are reversed, leaving $\tau$ unchanged. We can therefor drop the subscripts, and write:

\begin{equation} \tau(x) = \frac{1}{2a} \log(\frac{z + t}{z - t}) 
\label{eq:unruh1.1} \end{equation}
	which is the 
Rindler time-coordinate, and covers only region U. The hypersurfaces 
$\Sigma_{\tau_0}$ are given by $t_{\tau_0}(z) = z \tanh(a \tau_0)$, as 
shown in figure 2. The radar distance $\rho(x)$ (which is positive 
by construction) is given by $\rho(x)  =  \frac{|\log(a^2 (z^2 - t^2))|}{2a} = 
\frac{|\log(a^2 u^2)|}{2a}$, where $u \equiv \sqrt{z^2 - t^2}$. In $(\tau,u)$ 
coordinates the metric takes the familiar form:

\begin{equation} d s^2 = a^2 u^2 d \tau^2 - d u^2  \notag \end{equation}
	Two uniformly accelerating observers, each with different $a$, will have the same hypersurfaces of simultaneity, but will `tick off' these hypersurfaces at different rates. The 
vector field $k^{\mu}$ is given by:
\begin{equation} k = z \frac{\partial}{\partial t} - t \frac{\partial}{\partial z} \hs{1} 
\mbox{or } = \frac{\partial}{\partial \tau} \mbox{ in Rindler 
coordinates} \notag \end{equation}
	which is the Killing vector field that is used to define positive/negative 
frequency modes in conventional derivations of the Unruh effect. 

\begin{figure}[h]
\center{\epsfig{figure=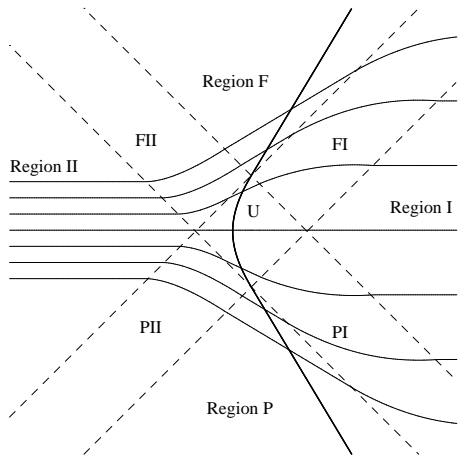, width=6cm}}
\caption{{\footnotesize Hypersurfaces of simultaneity of an observer 
undergoing uniform acceleration for a finite period of time (in Region 
U here) who is otherwise inertial.}}
\end{figure}

To see the 
significance of the dotted line in region II, consider the `Finite 
Acceleration Time' case shown in figure 3. In the 
limit as the `Acceleration Time' approaches infinity, this case approaches that 
of a uniformly accelerating observer. In this limit the hypersurfaces of simultaneity 
(which are Cauchy) all approach this dotted line in region II ($k_{\mu}(x) 
\rightarrow 0$ there), and a particle horizon forms. This explains the 
requirement (essential to many derivations of the Unruh Effect 
\cite{SMG,GMR,BD,BMPS}) that the `Rindler modes' must be those that are 
zero throughout region II. Further details of the `Finite 
Acceleration Time' case can be found in \cite{Me3}.

Consider now the 3+1 dimensional case, with $x^{\mu}_{(\gamma)}(\tau) =  
(\frac{\sinh(a \tau)}{a} , 0 , 0 , \frac{\cosh(a \tau)}{a})$. In this case 
equations (\ref{eq:deriv1}) can be replaced by (dropping subscripts)
\begin{equation} (t(\tau^+) - t)^2 = (z - z(\tau^+))^2 + |\bx_{\perp}|^2 \notag \end{equation}
(where $|\bx_{\perp}|^2 \equiv x^2 + y^2$), which applies to both $\tau^+$ and $\tau^-$. 
Substitution of the expressions for $t(\tau)$ and $z(\tau)$ and rearrangement gives:
\begin{equation} T_{\pm}^2 - \left( \frac{1}{a(z-t)} + a(z+t) \right) T_{\pm} + 
\frac{z+t}{z-t} = 0 \label{eq:5.1.5} \end{equation}
where $T_{\pm} \equiv e^{a \tau^{\pm}}$ are the two roots of (\ref{eq:5.1.5}). From 
this it follows immediately that $e^{2 a \tau} = T_+ T_- = \frac{z+t}{z-t}$, 
just as in the 1+1 dimensional case. $k_{\mu}(x)$ is also as before, and 
we can again define $u \equiv \sqrt{z^2 - t^2}$ so that the metric 
can be written as
\begin{equation} d s^2 = a^2 u^2 d \tau^2 - d u^2 - d x^2 - d y^2  \notag \end{equation}
 	reproducing the Rindler coordinates.

	We have now justified the use of Rindler coordinates for studying 
uniformly accelerating observers, and the derivation of the conventional Unruh 
effect becomes standard \cite{GMR,SMG,Tak}. We reproduce the massive and 
massless cases here, restricting ourselves to 1+1 dimensions for convenience.

	By defining $\gamma_0 = a u \bar{\gamma}_0$ and $\gamma_3 = \bar{\gamma}_3$
 (where $\bar{\gamma}_0, \bar{\gamma}_3$ are any matrices satisfying 
$\bar{\gamma}_0^2 = 1 = - \bar{\gamma}_3^2$ and $\{ \bar{\gamma}_0, 
\bar{\gamma}_3\} = 0$) we find that $\Gamma_0 = \half a \bar{\sigma}_3$ and 
$\Gamma_3 = 0$. The Dirac equation can be written in the form of (\ref{eq:dissHam2}) as:

\begin{equation} i \frac{\partial \psi}{\partial \tau} + \frac{i}{2} a 
\bar{\sigma}_3 = a u (-i \bar{\sigma}_3 \frac{\partial \psi}{\partial u}) + a u 
\bar{\gamma}^0 \psi \label{eq:unruh3} \end{equation}

The inner product in U is written in these coordinates as:
\begin{equation} \la \psi | \phi \ra = \int_0^{\infty} \psi^{\dagger} \phi du 
\label{eq:unruh3.1} \end{equation}

	The $u \frac{\partial}{\partial u}$ term 
on the RHS of (\ref{eq:unruh3}) implies, as expected, that $\hat{H}_{ev}$ is 
not Hermitian. A simple calculation reveals that $\hat{H}^{\dagger}_{ev} = \hat{H}_{ev} - 
i a \bar{\sigma}_3$. Hence (\ref{eq:unruh3}) can be rewritten in terms of $\hat{H}_1$ as:

\begin{equation} i \frac{\partial \psi}{\partial \tau}  = \hat{H}_1 \psi 
\equiv a u (-i \bar{\sigma}_3 \frac{\partial \psi}{\partial u}) + a u 
\bar{\gamma}^0 \psi - \frac{i}{2} a \bar{\sigma}_3 \label{eq:unruh4} \end{equation}
	It is easy to verify that (\ref{eq:Hsig2}) is indeed satisfied. The 
eigenstates of $\hat{H}_1$ are the stationary states, justifying again the 
choice made in previous derivations. Since the eigenstates of $\hat{H}_1$ are stationary, then although the Rindler 
observer disagrees that the Minkowski vacuum is empty, he still agrees that 
the particle content does not change with $\tau$. 

\subsubsection{Solutions in Region U}

	Consider solutions of the form:
$$ ^I\psi_{\omega}(x) = (f(u) \phi_+ + g(u) \phi_-) \psi_{\omega}(u) e^{-i \omega \tau } $$
where $\phi_{\pm}$ are basis spinors, satisfying
$$\phi^{\dagger}_{\lambda} \phi_{\sigma} = 2 \delta_{\lambda \sigma} \hs{.5}
\bar{\sigma}_3 \phi_{\pm} = \pm \phi{\pm} \hs{.5} \bar{\gamma}_0 \phi_{\pm} = 
\phi{\mp}$$ The factor of 2 is for convenient comparison with the 
representation given in \cite{GMR}. These are eigenstates of $\hat{H}_1$ for 
all $\tau$, with eigenvalue $\omega$.

	Substitution into (\ref{eq:unruh4}) gives:

\begin{align} \frac{i}{u} \left( u \frac{d}{d u} + \half -i \frac{\omega}{a} 
\right) f(u) & = m g(u) \notag \\
 \frac{-i}{u} \left( u \frac{d}{d u} + \half + i \frac{\omega}{a} 
\right) g(u) & = m f(u) \label{eq:CE6} \end{align} 

from which we find that
\begin{align} 
(u \frac{d}{d u} u \frac{d}{d u}) f(u)
& = \left[m^2 u^2 - (\frac{\omega}{a} + \frac{i}{2})^2 \right] f(u)
\label{eq:CE9.1} \\
(u \frac{d}{d u} u \frac{d}{d u}) g(u)
& = \left[m^2 u^2 - (\frac{\omega}{a} - \frac{i}{2})^2 \right] g(u)
 \label{eq:CE9.2} \end{align} 
Equations (\ref{eq:CE9.1}) and (\ref{eq:CE9.2}) can be identified with equation (21.66) of \cite{GMR}, or with 
page 4 of \cite{Bat}.
The only normalisable solution is of the form:
\begin{equation}
^I\psi_{\omega}(\tau,u) = \{ 
H^{(1)}_{\frac{i \omega}{a}  - \half}(i m u) \phi_{+} +   
H^{(1)}_{\frac{i \omega}{a} + \half}(i m u) \phi_{-} \} 
e^{-i w \tau} \notag  \end{equation}

where $H^{(1)}_{\nu}(z)$ are Hankel Functions. These states satisfy \cite{GMR}:
$$ \la ^I\psi_{\omega} | ^I\psi_{\omega'} \ra = \frac{16 a 
\delta(\omega - \omega')}{m (1 + e^{\frac{-2 \pi \omega}{a}})}$$

	We shall return to these solutions after finding a convenient 
representation for the Minkowski modes.

\subsection{The Minkowski modes}

	We have just found a basis for $\clh^{+}_R$ and 
$\clh^{-}_R$. Together they span all those modes that can be seen by 
the Rindler observer. They span only half of $\clh_M$, since they 
do not include any states defined in region II. However, they are 
sufficient to calculate the Rindler 
number operator $\hat{N}_{1,R} = \hat{P}^+_R - \hat{P}^-_R$, from  
which we can deduce the number of Rindler particles present in the 
Minkowski vacuum, along with their frequency and spatial distributions. To 
define the Minkowski vacuum, we 
could use the ordinary plane wave basis for Minkowski 
modes. However, it is more convenient for the massive case to use 
an alternative representation of Minkowski modes \cite{GMR,Tak,SMG}. As a consistency check, we shall rederive 
the massless limit using Minkowski plane wave states. 

The representation of Minkowski modes used in \cite{GMR} (often called 
the `Rindler Basis') is of the form:

\begin{align}
u^{M}_{\omega} & = N ( {}^I\psi_{\omega} - e^{\frac{\pi \omega}{2 a}} 
{}^F \psi_{\omega}^{(+)} + e^{\frac{- \pi \omega}{2 a}} {}^P \psi_{\omega}^{(+)} 
- i e^{\frac{- \pi \omega}{a}} {}^{II}\psi_{\omega} ) \label{eq:unruh6} \\
v^{M}_{\omega} & = N (e^{\frac{- \pi \omega}{a}} {}^I\psi_{\omega} + 
e^{\frac{- \pi \omega}{2 a}} {}^F \psi_{\omega}^{(-)} + e^{\frac{\pi \omega}{2 a}} 
{}^P \psi_{\omega}^{(-)} + i {}^{II}\psi_{\omega} ) \label{eq:unruh7} \end{align}

where $^{II}\psi_{\omega}(x) = \bar{\sigma}_3 ^{I}\psi_{\omega}(-x)$ 
is a state defined in Region II, and $^F\psi_{\omega}^{(\pm)}, 
^P\psi_{\omega}^{(\pm)}$ are states defined in Regions F and P 
respectively (see \cite{GMR,SMG} for details). These various 
states are defined only in their particular region, so the 
Dirac operator acting on any of these states does not give zero, 
but gives a distribution on the light cone through the 
origin. The coefficients in the above expansion are 
chosen to cancel these distributions. The $u^{M}_{\omega}(x)$ then form an orthogonal basis for 
$\clh^{+}_M$, and the $v^{M}_{\omega}(x)$ form an 
orthogonal basis for $\clh^{-}_M$ (irrespective of the 
sign of $\omega$).

	In (\ref{eq:unruh6}) and (\ref{eq:unruh7}), the states 
$^{II}\psi_{\omega}, ^{I}\psi_{\omega},^F\psi_{\omega}^{(\pm)}, 
^P\psi_{\omega}^{(\pm)}$ are not normalised. By rewriting (\ref{eq:unruh6}) 
and (\ref{eq:unruh7}) in terms of states $^{II}\psi_{\omega}, ^{I}\psi_{\omega}$ 
which are each normalised to $2 \pi \delta(\omega - \omega')$, and choosing 
$N$ such that $u^{M}_{\omega}(x)$ and $v^{M}_{\omega}(x)$ are normalised 
to $2 \pi \delta(\omega - \omega')$, equations (\ref{eq:unruh6}) and 
(\ref{eq:unruh7}) become:

\begin{align}
u^{M}_{\omega} & = \frac{1}{\sqrt{1 + e^{\frac{- 2 \pi \omega}{a}}}} ( 
{}^I\psi^{\rm{norm}}_{\omega}  - i e^{- \pi \omega} {}^{II}\psi^{\rm{norm}}_{\omega} 
+ \mbox{ F, P, terms} ) \notag \\
v^{M}_{\omega} & = \frac{1}{\sqrt{1 + e^{\frac{- 2 \pi \omega}{a}}}} (e^{- \pi \omega} 
{}^I\psi^{\rm{norm}}_{\omega}  + i  {}^{II}\psi^{\rm{norm}}_{\omega} + 
\mbox{ F, P, terms} ) \notag \end{align}
	from which we can immediately extract the relations:

\begin{align}
\alpha^I_{\omega \omega'} & \equiv \la {}^I\psi_{\omega} | u^M_{\omega'} \ra 
= \frac{2 \pi \delta(\omega - \omega')}{\sqrt{1 + e^{\frac{-2 \pi \omega}{a}}}} 
\label{eq:unruh8} \\
\beta^I_{\omega \omega'} & \equiv \la {}^I\psi_{\omega} | v^M_{\omega'} \ra 
= \frac{2 \pi \delta(\omega - \omega')}{\sqrt{1 + e^{\frac{2 \pi \omega}{a}}}} \\
\gamma^I_{\omega \omega'} & \equiv \la {}^I\psi_{-\omega} | u^M_{\omega'} \ra 
= \frac{2 \pi \delta(\omega + \omega')}{\sqrt{1 + e^{\frac{2 \pi \omega}{a}}}}  \\
\epsilon^I_{\omega \omega'} & \equiv \la {}^I\psi_{-\omega} | v^M_{\omega'} \ra
= \frac{2 \pi \delta(\omega + \omega')}{\sqrt{1 + e^{\frac{-2 \pi \omega}{a}}}} 
\label{eq:unruh9} \end{align}
	where $\omega$ is restricted to $\omega > 0$, but $\omega'$ 
can take any sign. Alternatively, these coefficients can be extracted 
using the observation that $\frac{u_{\omega}^M}{\sqrt{1 + 
e^{\frac{-2 \pi \omega}{a}}}} + \frac{v_{\omega}^M}{\sqrt{1 + 
e^{\frac{2 \pi \omega}{a}}}}$ (equal to ${}^I\psi^{\rm{norm}}_{\omega} + 
\mbox{ F, P, terms}$ ) is the only linear combination of $u_{\omega}^M$ 
and $v_{\omega}^M$ that is zero in Region II. Although 
the inner products in (\ref{eq:unruh8}) - (\ref{eq:unruh9}) have 
been evaluated on a hypersurface (from (\ref{eq:unruh3.1})) which 
is not Cauchy in Minkowski space (it covers space according to the 
Rindler observer), it is still consistent with any inner product in 
Minkowski space, since the Rindler modes are all zero in region II.

	From (\ref{eq:unruh8}) - (\ref{eq:unruh9}) it follows that: 
\begin{equation} N^{\pm}_{\omega} = \rm{Trace}(\beta \beta^{\dagger}) = 
\int_{-\infty}^{\infty} \frac{d \omega'}{2 \pi} \frac{2 \pi \delta(\omega 
\pm \omega')}{1 + e^{\frac{2 \pi \omega}{a}}} = \frac{1}{1 + 
e^{\frac{2 \pi \omega}{a}}} \label{eq:N1} \end{equation}
	which represents a thermal spectrum at Temperature $T = 
\frac{a}{2 \pi k_B}$, as expected.

\subsubsection{The Massless Case}

	In the massless case equations (\ref{eq:CE6}) decouple, and we find 
two independent solutions for each $\omega$,

\begin{equation} | \psi_{\omega,1} \ra = \phi_+ (a u)^{\frac{i \omega}{a} - \half}
 e^{-i \omega \tau} \mbox{ and } | \psi_{\omega,2} \ra = \phi_- (a u)^{\frac{-i 
\omega}{a} - \half} e^{-i \omega \tau} \label{eq:unruh10} \end{equation}
Each of these is normalised to $\la \psi_{\omega,\sigma} |\psi_{\omega',\sigma'} 
\ra = (2 \pi) \delta_{\sigma \sigma'} \delta(\omega - \omega')$. The 
massless limit of $| ^{I}\psi_{\omega}^{\rm{norm}} \ra$ would yield one 
specific linear combination of $| \psi_{\omega,1} \ra$ and 
$| \psi_{\omega,2} \ra$. However, we shall find that $| \psi_{\omega,1} \ra$ and 
$| \psi_{\omega,2} \ra$ each lead to a thermal spectrum, so that we need not  
restrict ourselves to this one linear combination.

	The plane wave states $\clh^{\pm}_M$ also take a particularly simple form 
in the massless case. A basis of $\clh^+_M$ is provided by states of the form:
$$ \phi_+ e^{- i p (t-x)} \mbox{ and } \phi_- e^{ -i p (t+x)} \mbox{ for } p>0 $$
 and a basis for $\clh^-_M$ is provided by states of the form:
$$ \phi_+ e^{i p (t-x)} \mbox{ and } \phi_- e^{i p (t+x)} \mbox{ for } p>0 $$
	This allows us to write:
\begin{align} N^+_{\omega,1} & = \sum_p |\beta_{\omega,1  p}|^2 = \frac{1}{L} 
\int_0^{\infty} \frac{d p}{2 \pi} | \la \psi_{\omega,1} | \phi_+ 
e^{ i p(t - x)} \ra|^2 \notag \\
\mbox{where } \la \psi_{\omega,1} | \phi_+ e^{ i p(t - x)} \ra & = 
\int_0^{\infty} d x e^{- i p x} (a x)^{\frac{-i \omega}{a} - \half} \hs{.2} = 
\frac{1}{a} (\frac{p}{a}e^{\frac{i \pi}{2}})^{\frac{i \omega}{a} - \half} 
\Gamma(\frac{-i \omega}{a} + \half) \notag \end{align}
	and $L \equiv \la \psi_{\omega,1} |\psi_{\omega,1} \ra = 
\int_0^{\infty} \frac{d u}{a u} = (2 \pi) \delta(0)$. We have evaluated 
the inner product on the hypersurface $t=0$, and we have used \cite{Bat} 
(equation (6) page 1) in the last line. By using further properties of the 
$\Gamma$-function we deduce that:
\begin{equation} N^+_{\omega,1} = \frac{1}{L} \int_0^{\infty} \frac{d p}{a p} 
\frac{1}{1 + e^{\frac{2 \pi \omega}{a}}} = \frac{1}{1 + 
e^{\frac{2 \pi \omega}{a}}} \label{eq:Nn2} \end{equation}
as expected. Similarly, 
$$N^-_{\omega,1} = \int_0^{\infty} \frac{d p}{2 \pi} 
| \la \psi_{-\omega,1} | \phi_+ e^{- i p(t - x)} \ra|^2 = \frac{1}{1 + 
e^{\frac{2 \pi \omega}{a}}}$$ and it is straightforward to show that 
each of $N^{\pm}_{\omega,2}$ is also each equal this value.

\subsection{Spatial Distribution of Rindler Particles}

 Although (\ref{eq:N1}) and (\ref{eq:Nn2}) each show that the spectrum of 
Rindler particles is thermal, they do not tell us the spatial 
distribution of these particles. This can be deduced from (\ref{eq:numdens-}) and (\ref{eq:numdens+}). For 
this purpose, it is clearest  
to work with the radar-like spatial coordinate $\chi = \frac{1}{a} 
\log(a u)$. In these coordinates our observer is at $\chi = 0$, and 
$|\chi|$ represents the radar distance from the observer to the point 
$(\tau,\chi)$ (for any $\tau$). In these coordinates the inner 
product (\ref{eq:unruh3.1}) takes the form $\la \psi | \phi \ra = 
\int_{-\infty}^{\infty} e^{a \chi} \psi^{\dagger} \phi d \chi$. The massless 
states of equation (\ref{eq:unruh10}) are of the form 
$e^{\frac{-a \chi}{2}} \phi_{\pm} e^{-i \omega(\tau \mp \chi)}$. They 
are plane wave states in these coordinates, with the 
conformal factor $e^{\frac{-a \chi}{2}}$ cancelling the $e^{a \chi}$ 
in the inner product. Define the particle/antiparticle densities $n^{\pm}(\tau,\chi)$ 
such that $n^{\pm}(\tau,\chi) d \chi$ represents the total number of 
particles/antiparticles within $d \chi$ of $\chi$. From 
(\ref{eq:numdens-}) and (\ref{eq:numdens+}) this is given by:
\begin{align}
n^{\pm}(\tau,\chi) & = \int_0^{\infty} \frac{n_{\pm \omega}(\tau,\chi)}{1 + e^{\frac{2 \pi \omega}{a}}} 
\frac{{\rm d} \omega}{2 \pi} \label{eq:unruh11} \\
\mbox{ where } \hs{1} n_{\omega}(\tau,\chi) & = e^{a \chi} {}^I\psi_{\omega}^{\dagger 
\rm{norm}}(\tau,\chi) {}^I\psi^{\rm{norm}}_{\omega}(\tau,\chi) \notag \end{align}

	Both are independent of $\tau$. 

	For the massless case, $\psi_{\omega}^{\dagger} \psi_{\omega} = 
e^{-a \chi}$, so that $n^+(\tau,\chi) = n^-(\tau,\chi)$ and both are spatially 
uniform in $\chi$. However, as mentioned in Section 3.5, the 
interpretation of $n^{\pm}(\tau,\chi)$ as representing 
`particle density' is only accurate when averaged over a distance $L$ 
sufficient to suppress the fluctuations in $\hat{N}^{\pm}_L(\tau)$.
From (\ref{eq:fluc80}) and (\ref{eq:fluc90})), a straightforward calculation yields:

\begin{align} \la (\hat{N}^+_L)^2 \ra & = \la (\hat{N}^-_L)^2 \ra \notag \\
& = \int_0^{\infty} \int_0^{\infty} \frac{d \omega}{2 \pi} \frac{d \omega'}{2 \pi}
\frac{\sin^2((\omega - \omega')L)}{(\omega - \omega')^2} \frac{e^{\frac{2 \pi \omega'}{a}}}{1 + e^{\frac{2 \pi \omega'}{a}}} \frac{1}{1 + e^{\frac{2 \pi \omega}{a}}} \notag \\
 & = \frac{a^2 L^2}{4 \pi^2} \int_0^{\infty} {\rm d} \mu \frac{\sin^2(a \mu L)}{\mu^2 a^2 L^2} \cosh(\pi \mu) L(\mu) \label{eq:90} \\
\mbox{where } \hs{1} L(\mu) & \equiv \int_{|\mu|}^{\infty} \frac{d \lambda}{\cosh(\pi \mu) + \cosh(\pi \lambda)} \notag \\
& = \frac{- \log[\cosh(\pi \mu) (\sqrt{\cosh^2(\pi \mu) + 1} - 1)]}{\pi \sqrt{\cosh^2(\pi \mu) + 1}} \notag \end{align}

\begin{figure}[hbt]
\begin{minipage}[b]{6cm}
\epsfig{figure=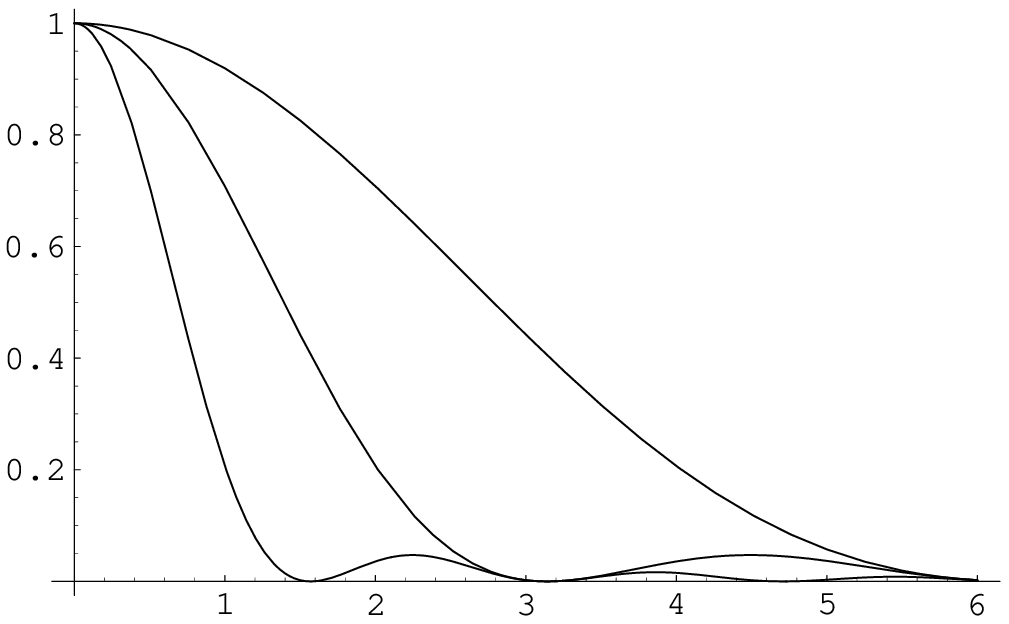,width=6cm}

{\footnotesize {\bf FIG. 4 A.} $\frac{\sin^2(a \mu L)}{\mu^2 a^2 L^2}$ as a function of $\mu$, for $L = \frac{2}{a}$ (bottom curve) $\frac{1}{a}$ and $\frac{1}{2 a}$ (top curve).}
\end{minipage}\hs{.5}\begin{minipage}[b]{5cm}
\epsfig{figure=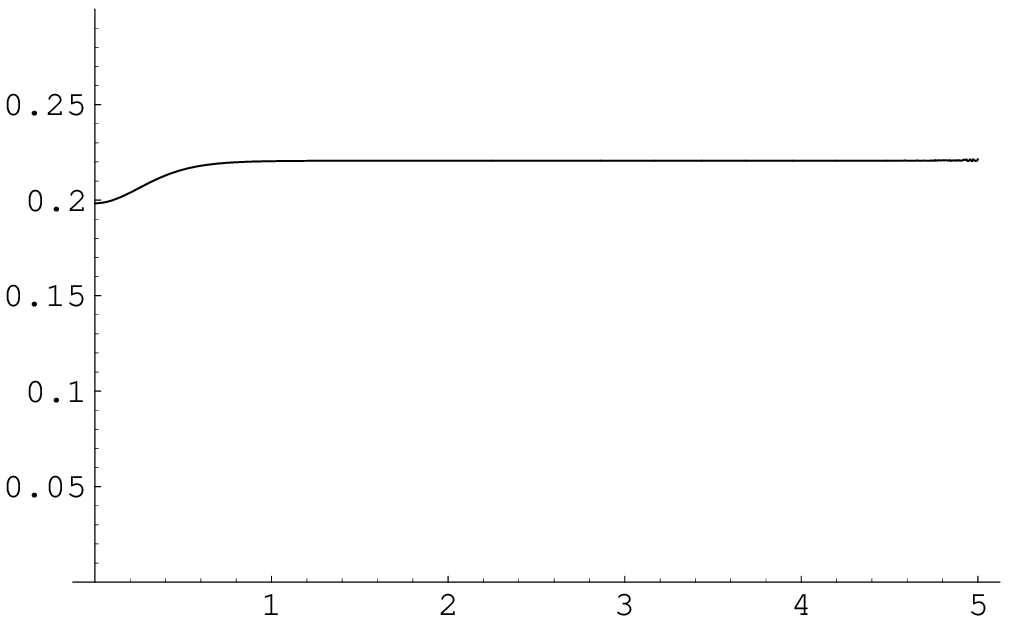,width=5cm}

{\footnotesize {\bf FIG 4 B.} $\cosh(\pi \mu) L(\mu)$ as a function of $\mu$. This function is $\frac{\log(\sqrt{2} + 1)}{\pi \sqrt{2}}$ at $\mu=0$, but quickly approaches its large $\mu$ limit of $\frac{\log(2)}{\pi}$.}
\end{minipage}

\end{figure}

\addtocounter{figure}{1}

Figure 4 B shows $\cosh(\pi \mu) L(\mu)$ as a function of $\mu$, while 
Figure 4 A shows $\frac{\sin^2(a \mu L)}{\mu^2 a^2 L^2}$ for $L = 
\frac{2}{a}$ (bottom curve) $\frac{1}{a}$ and $\frac{1}{2 a}$ (top curve).
For $L >> \frac{1}{a}$ the integral in (\ref{eq:90}) is dominated by 
contributions from small $\mu$. We can then approximate $\cosh(\pi \mu) 
L(\mu) \approx \frac{\log(\sqrt{2} + 1)}{\pi \sqrt{2}} \approx .198$. In this limit:
\begin{equation} \la (\hat{N}^+_L)^2 \ra  \approx a L \frac{\log(\sqrt{2} + 1)}{8 
\pi^2 \sqrt{2}} \approx .45 N^+_L \end{equation}
	For $L >> \frac{1}{a}$ we can approximate $\cosh(\pi \mu) L(\mu) 
\approx \frac{\log(2)}{\pi}$, so that:
\begin{equation} \la (\hat{N}^+_L)^2 \ra  \approx a L \frac{\log(2)}{8 \pi^2} = 
.5 N^+_L \end{equation}
	In either case, we confirm that:
\begin{itemize} 

\item $ \la (\frac{\hat{N}^+_L}{L})^2 \ra \approx \frac{1}{2 L} \frac{N^+_L}{L} \propto \frac{1}{L}$, which would be infinite for $L \rightarrow 0$, but is finite other wise.  

\item The uncertainty in the number of particles in length L, given by $\sqrt{\la (\hat{N}^+_L)^2 \ra}$, is proportional to the square root of the number of particles in that length. For this uncertainty to be less than the measurement, we must average over a length such that $N^+_L > 1$. This is a reasonable requirement - if we want a reliably defined particle density, we must average over a volume that contains on average more than 1 particle.

\end{itemize}

Now we study the massive case. We can show from properties of the Hankel 
functions \cite{Bat} that $n_{\omega}(\tau,\chi) = n_{-\omega}(\tau,\chi)$, 
so that the `principle of detailed balance' \cite{Tak} applies at every  
point in Rindler space. The relation $n_{\omega}(\tau,\chi) = n_{-\omega}(\tau,\chi)$ also 
expresses the fact that the distribution of Rindler antiparticles 
exactly matches the distribution of Rindler particles.

\begin{figure}[h]
\center{\epsfig{figure=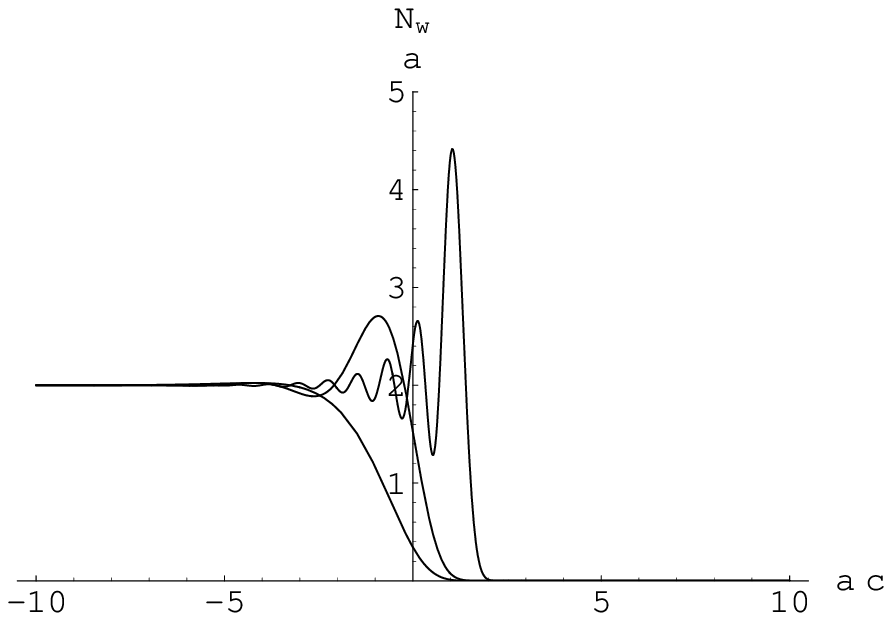, width=8cm}}
\caption{{\footnotesize $N_{\omega}(\tau,\chi)/a$ as a function of $a \chi$, for $m=a$, 
and $\omega = \frac{a}{4}$ (lowest curve), $a$ and $4 a$ (most oscillatory curve).}}
\end{figure}

	In Figure 5 we have plotted $n_{\omega}(\tau,\chi)/a$ as a 
function of $a \chi$, for $m=a$ and $\omega = \frac{a}{4}$, $a$ 
and $4 a$. With $c$ and $\hbar$ included, a length $a \triangle \chi = 1$ corresponds to 
$\triangle \chi = \frac{c^2}{a}$. Masses are measured in units of 
$\frac{a \hbar}{c^3}$ and frequencies in units of $\frac{a}{c}$. Normalisation of the massless 
plane wave states to $(2 \pi) \delta(\omega - \omega')$ represents a 
norm of `1 per unit length', as is conventional. For massive states this 
interpretation is valid only as a Cauchy 
principal value, so that $\frac{1}{L} \int_{-\frac{L}{2}}^{\frac{L}{2}} 
N_{\omega}(\tau,\chi) d \chi \rightarrow 1$ as $L \rightarrow \infty$.)

Figure 5 shows that for negative $\chi$ the Rindler particles are 
uniformly distributed, but for positive $\chi$ (i.e. to the 
observer's right) the number of particles decreases 
rapidly. This fact can be understood \cite{Tak} by writing 
equation (\ref{eq:CE9.1}) in terms of $\chi$ as:

\begin{equation} - \frac{d^2 f}{d \chi^2} + m^2 e^{2 a \chi} f = (\omega + 
\frac{i a}{2})^2 f \label{eq:potential} \end{equation}
which takes the form of a 1-dimensional Schr\"{o}dinger equation with 
potential $V = m^2 e^{2 a \chi}$. For $m=0$ the potential disappears and 
we obtain the plane wave states described above. For $a = 0$ the potential 
allows $V = m^2$ and the mass gap reappears. For nonzero $m, a$ the 
potential permits states of any frequency, but does not allow 
them to propagate further to the right than $\chi = \frac{1}{a} 
\log(\frac{\omega}{m})$ before being exponentially damped. We also see 
from the form of (\ref{eq:potential}) that a change in mass by a factor 
$e^{a l}$ has the effect of translating the potential a distance 
$l$ to the left. This is illustrated in Figure 6.

\begin{figure}[h]
\center{\epsfig{figure=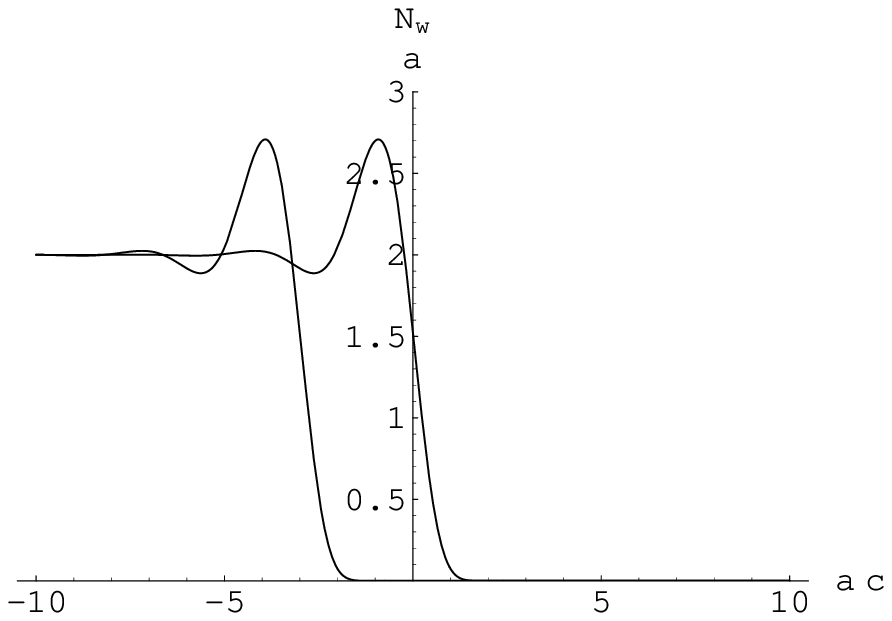, width=8cm}}
\caption{{\footnotesize $N_{\omega}(\tau,\chi)/a$ as a function of $a \chi$, 
for $w=a$, and $m = a$ (right curve)and $a e^3$ (left curve).}}
\end{figure}

	Figure 7 (A) depicts $n(\tau,\chi)/a$ as a function of $a \chi$, for 
$m = a/10$ (right curve), $a$, and $10 a$, obtained by 
integrating (\ref{eq:unruh11}) numerically over $0 \leq \omega \leq 2 a$. The factor 
$\frac{1}{1 + e^{\frac{2 \pi \omega}{a}}}$ ensures that the number 
of particles present having energy $>2 a$ is negligible (see Fig. 8(B)). We would 
like to interpret $n(\tau,\chi)/a$ as the number density at $\chi$, and 
$\frac{n_{\omega}(\tau,\chi)}{a(1 + e^{\frac{2 \pi \omega}{a}})}$ 
as `the number density at $\chi$ 
of particles of frequency $\omega$'. This again requires caveats regarding 
fluctuations, in the same way as in the massless 
case. Accordingly, we find   
that the particle number density is uniform to the observer's left and 
negligible to the observer's right, and that the position of this transition 
is determined by the ratio $\frac{m}{a}$.

For $m \rightarrow 0$ this 
transition point goes to $\infty$, reproducing the spatial uniformity 
of the massless limit. However, for non-zero $m$ and realistic accelerations, 
the particle density at low $\chi$ (which is $\propto a$) 
becomes small, and the transition to a negligible density occurs far to the 
observer's left.

\begin{figure}[hbt]
\begin{minipage}[b]{5.5cm}
\epsfig{figure=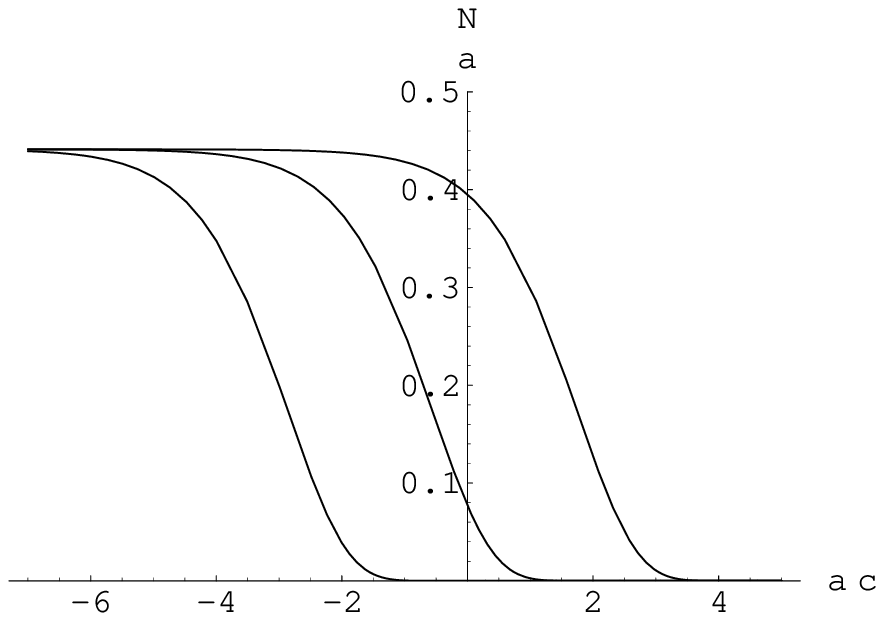,width=5.5cm}

{\footnotesize {\bf FIG. 7 A.} $N(\tau,\chi)/a$ as a function of $a \chi$, for $m = a/10$ (right curve), $a$, and $10 a$ (left curve).}
\end{minipage}\hs{.5}\begin{minipage}[b]{5.5cm}
\epsfig{figure=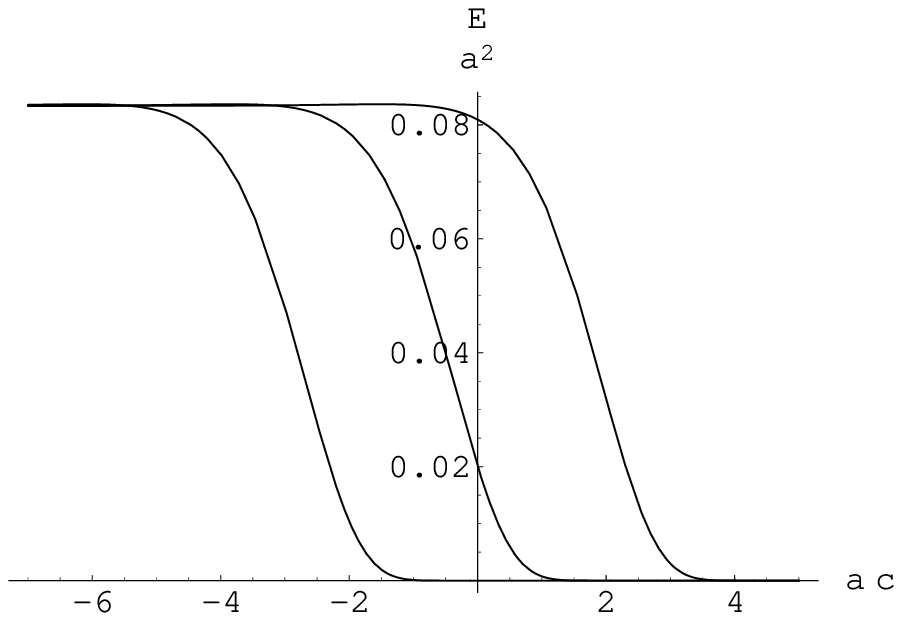,width=5.5cm}

{\footnotesize {\bf FIG 7 B.} $E(\chi)/a^2$ as a function of $a \chi$, for $m = a/10$ (right curve), $a$, and $10 a$ (left curve).}
\end{minipage}

\end{figure}

\addtocounter{figure}{1}

Consider now the energy density of these Rindler particles as measured by the 
Rindler observer. It is given by $E(\chi) = E^+(\chi) + E^-(\chi)$ where 
$$E^{\pm}(\chi) = \int_{0}^{\infty} \frac{\omega n_{\pm \omega}(\tau,\chi)}{1 
+ e^{\frac{2 \pi \omega}{a}}} \frac{{\rm d} \omega}{2 \pi}$$ 
Since $n_{\omega}(\tau,\chi)$ is even in $\omega$ and is independent of $\tau$, 
it follows that $E^+(\chi) = E^-(\chi)$, and is independent of $\tau$ as expected. 
Figure 7 (B) shows $\frac{E(\chi)}{a^2}$ as a function of $a \chi$, for 
$m = a/10$ (right curve), $a$, and $10 a$. This is qualitatively the same as 
Figure 7 (A), as expected.

\begin{figure}[hbt]
\begin{minipage}[b]{6.5cm}
\epsfig{figure=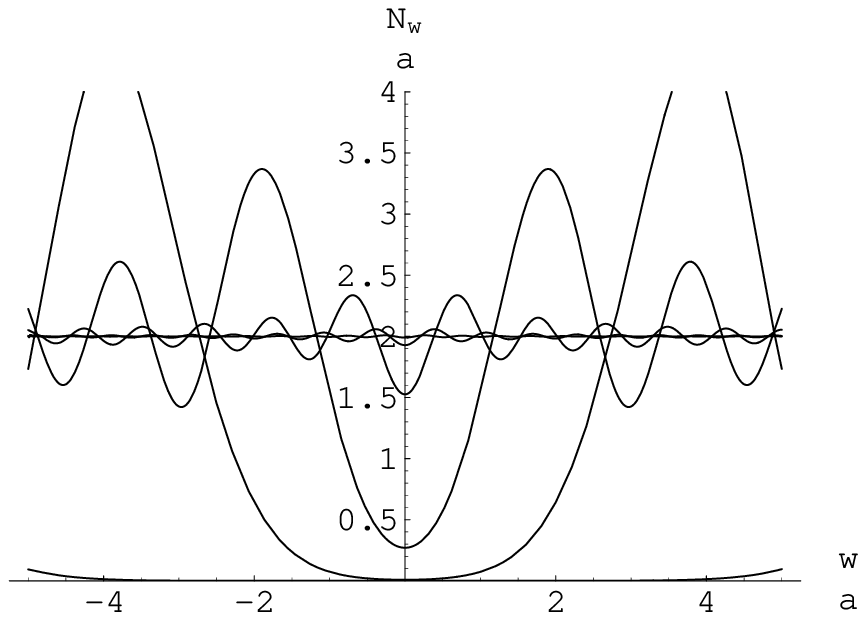,width=7cm}

{\footnotesize {\bf FIG. 8 A.} $N_{\omega}(\tau,\chi)/a$ as a function of 
$\frac{\omega}{a}$ for $m=a$ and $a \chi = -6,-4,-2,0,1$ and $2$}
\end{minipage}\hs{.5}\begin{minipage}[b]{4cm}
\epsfig{figure=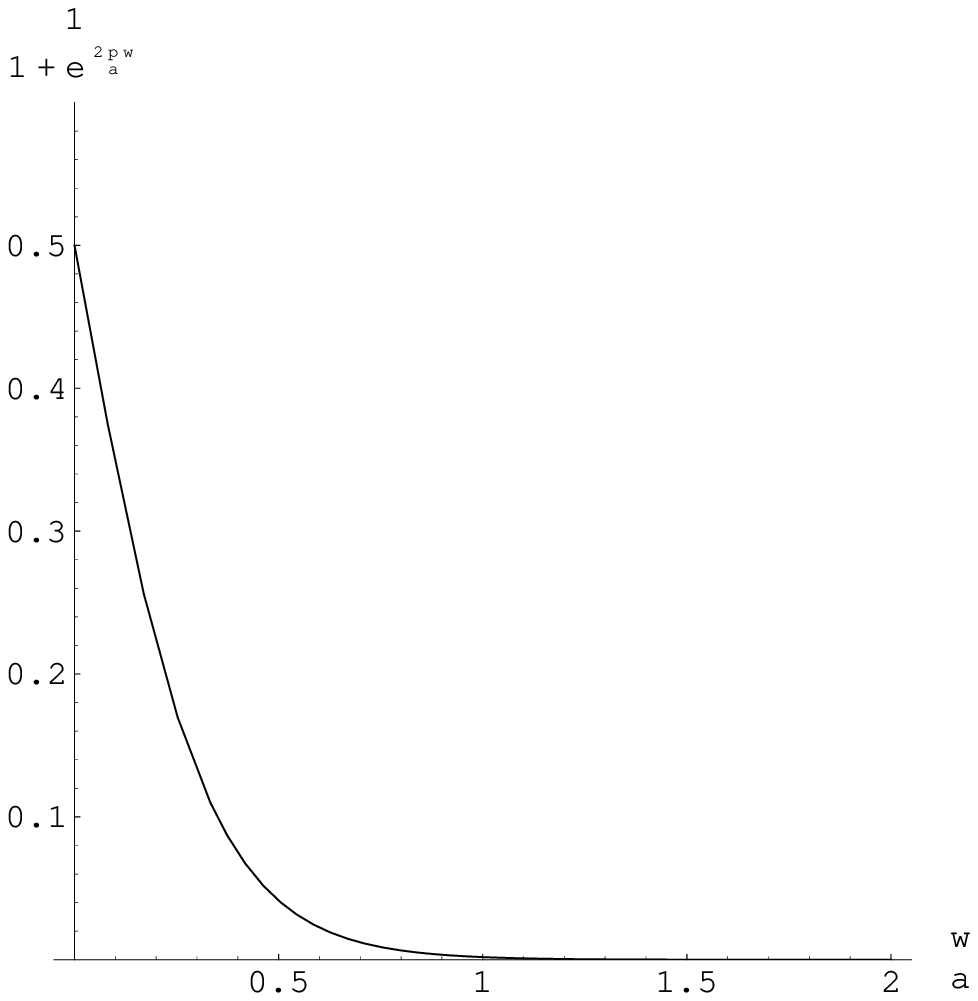,width=4cm}

{\footnotesize {\bf FIG 8 B.} $(1 + e^{\frac{2 \pi \omega}{a}})^{-1}$ as 
a function of $\frac{\omega}{a}$.}
\end{minipage}

\end{figure}

\addtocounter{figure}{1}

	It is not surprising that the Rindler observer detects the 
presence of particles and of energy; this is fully 
consistent with detector models. It is changes in the energy 
levels of the detector that signal the detection of Rindler 
particles. However, we should not identify the `energy 
as measured by an accelerating observer' with the covariant energy momentum 
tensor $\la T_{\mu \nu} \ra$, which is zero in the Minkowski vacuum, albeit 
with non-zero fluctuations. To understand why there is no conflict 
between $E(\chi) \neq 0$ and $\la T_{\mu \nu} \ra = 0$, it is instructive to 
recall the classical connection between $T^{\mu \nu}$ and 
$E$. For instance (\cite{MTW}, pg 131) if we 
contract $T^{\mu \nu}$ with a 4-velocity $u_{\mu}$, defined at at some 
event $x$, then $T^{\mu \nu} u_{\nu}$ represents the ``4-momentum per 
unit of three-dimensional volume, `$\frac{d p^{\mu}}{d V}$', as 
measured in the observer's Lorentz frame at event $x$''. This interpretation 
necessarily involves division by a small 3-volume $d V$, which 
must be sufficiently small that the observer's Lorentz frame remains a 
good description of his `rest frame' (and also that $T^{\mu \nu}$ be 
effectively constant across $V$). This in turn requires the volume 
to have dimensions $\triangle x << c^2 a^{-1}$ where $a$ is the 
observers acceleration (\cite{MTW}, pg 169). But 
the uncertainty principle requires that $\triangle x > \frac{\hbar}{m c}$ 
(for the uncertainty in the momentum to be less than $m c$). Hence, the 
connection between $T^{\mu \nu} u_{\nu}$ and the ``4-momentum as measured 
by an observer'' applies only when $a << \frac{m c^3}{\hbar}$ - it breaks 
down precisely when the Unruh effect becomes significant. 

	We have not discussed either $\la T_{\mu \nu} \ra$ 
or the backreaction equation $G_{\mu \nu} = 8 \pi \la T_{\mu \nu} \ra$ 
which effectively defines it. These issues are discussed extensively in the 
literature (see for example \cite{BD,DeWitt,Full} and references therein).
We do not add to this discussion, except to emphasise that 
there is no conflict between that literature and the observer-dependence 
discussed here. Though the Hermitian operators representing `observables' 
may be non-local and observer-dependent, these do not affect 
the evolution of states, which is still local and causal \cite{Me1,mythesis}. There is no 
confict between seeking observer-dependent `observables' and seeking an 
observer-independent semiclassical evolution equation.

	Figure 8 (A) depicts $n_{\omega}(\tau,\chi)/a$ as a function of 
$\frac{\omega}{a}$ for $m=a$ and $a \chi = -6,-4,-2,0,1$ and $2$. We have 
also shown $\omega<0$ on this plot, to demonstrate that $n_{\omega}(\tau,\chi)/a = 
n_{-\omega}(\tau,\chi)/a$ as claimed earlier. This detailed balance  
also implies that the distribution of Rindler antiparticles exactly 
matches the distribution of Rindler particles. For $\chi = \frac{-6}{a}$, 
where the Rindler particles are uniformly distributed, 
$n_{\omega}(\tau,\chi)/a$ is completely flat, so that the 
spectrum of particles is exactly thermal there. At $\chi = \frac{-4}{a}$ 
we see slight deviations from a thermal spectrum, becoming more 
pronounced as $\chi$ increases. At $\chi = \frac{1}{a}$ the deviations from thermal are such 
that the low frequency modes are greatly suppressed. The factor of $(1 + e^{\frac{2 \pi \omega}{a}})^{-1}$ in 
(\ref{eq:unruh11}) (plotted in FIG. 8 (B)) suppresses all modes of 
frequency $\omega > a$, so that overall there are very few particles present at 
$\chi = \frac{1}{a}$, in accord with the middle curve of Figures 7 (A) and 7 (B).

\section{Further Examples of Radar Time}

	The radar time of uniformly accelerating observers in 1+1 or 
3+1 dimensions has been presented here and elsewhere 
\cite{Me3,mythesis,PaVa,LR}. A larger family of observers in 1+1 dimensions, 
including the `instant turnaround twin' (Langevin observer), and 
the `gradual turn-around' twin of Figure 3 have been described in detail in 
\cite{Me3}. In this Section we will further generalise these results. In Section 
5.1 we present the radar time and radar distance of an arbitrary observer 
in 1+1 dimensional flat space and investigate some of their properties.  
Section 5.2 extends this to arbitrary observers in an arbitrary 1+1 dimensional 
spacetime. Sections 5.3 and 5.4 treat cosmological applications. We shall present 
the radar time of a comoving observer in an FRW universe of arbitrary 
scale factor $a(t)$ in 1+1 and in 3+1 dimensions, and examine in more detail some 
examples, including the Milne and deSitter 
universes (for the latter see also \cite{mythesis}). 

\subsection{1+1 Dimensional Flat Space}

	The path of an arbitrary observer is completely characterised 
by a non-decreasing function $z_+^{(\gamma)}(z_-)$, where $ z_{\pm} = t 
\pm z$ are null coordinates, and the proper time is given by $d \tau^2 = 
d z_+ d z_- = \frac{d z_+^{(\gamma)}}{d z_-} d z_-^2$ on the curve. Define 
the functions $k_{\pm}(z_{\pm})$ and 
$\tau^{\pm}(z_{\pm})$ by:
\begin{align} k_-(z_-) \equiv \sqrt{\frac{d z_+^{(\gamma)}(z_-)}{d z_-}} & \hs{1}  
k_+(z_+) \equiv \sqrt{\frac{d z_-^{(\gamma)}(z_+)}{d z_+}} = 
\frac{1}{k_-(z_-^{(\gamma)}(z_+))} \label{eq:6.1} \\
 \tau^+(z_+) \equiv \int_0^{z_+} k_+(u) d u & \hs{1} \tau^-(z_-) 
\equiv \int_0^{z_-} k_-(u) d u \label{eq:6.2} \end{align}

	The origin is chosen to coincide with the point 
$\tau=0$ on the curve. On the observer's trajectory $k_- d z_- = k_+ d z_+$ and 
$\tau^+ = \tau^- =$ proper time. However, these functions are also
defined off the trajectory. For points to the right of the observer, 
$\tau^{\pm}$ are as defined in Section 3, and for points to the left 
of the observer the roles of $\tau^{\pm}$ are reversed for later 
convenience. We can write 
the radar time and radar distance as:

\begin{align} \tau(z_+,z_-) & = \half (\tau^+(z_+) + \tau^-(z_-)) = \half 
\left( \int_0^{z_+} k_+(u) d u + \int_0^{z_-} k_-(u) d u \right) \notag \\
\rho(z_+,z_-) & = \half | \tau^+(z_+) - \tau^-(z_-) | = \half \left| 
\int_0^{z_+} k_+(u) d u - \int_0^{z_-} k_-(u) d u \right| 
\notag \end{align}

	To make this construction more concrete, consider the 
case of an inertial observer. In this case $z_+^{(\gamma)}(z_-) = k_-^2 z_-$ 
where $k_{\pm} = \sqrt{\frac{1 \mp v}{1 \pm v}} =$ constant. $k_-$ is 
the ``$k$'' of Bondi's `$k$-calculus'. $\tau^{\pm} = k_{\pm} z_{\pm}$, so that the radar time is 
$$\tau = \half \sqrt{\frac{1-v}{1+v}} z_+ + \half \sqrt{\frac{1+v}{1-v}} z_- 
\hs{.5} = \frac{t - v x}{\sqrt{1 - v^2}}$$ which is the time coordinate 
of the observer's rest frame, as expected.

	For a uniformly accelerating observer we have $z_+^{(\gamma)}(z_-) 
= \frac{z_-}{1 - a z_-}$, where the curve has been translated relative to 
Section 5 so that it passes through the origin. This gives $k_{\pm}(z_{\pm}) = 
\frac{1}{1 \pm a z_{\pm}}$, so that:
\begin{align} \tau & = \frac{1}{2 a} \log(\frac{1 + a z_+}{1 - a z_-}) = 
\frac{1}{2 a} \log(\frac{x+\frac{1}{a} + t}{x+\frac{1}{a} - t}) \notag \\
\rho & = \frac{1}{2 a} |\log((1 + a z_+)(1 - a z_-))| \notag \end{align}
	which are the translated versions of (\ref{eq:unruh1.1}), as expected.

	Returning to the general case, we can write the metric 
in coordinates $(\tau,\rho)$ as:
\begin{equation} d s^2 = \frac{d \tau^2 - d \rho^2}{k_+(z_+) k_-(z_-)} 
\notag \end{equation}
	To write$k_{\pm}(z_{\pm})$ in terms of $\tau_{\pm}$ involves  
inverting the expressions (\ref{eq:6.2}) for $\tau_{\pm}(z_{\pm})$, which 
is always possible since $\tau_{\pm}(z_{\pm})$ are both strictly increasing 
functions (for a future-directed timelike observer). It is easy to verify 
that $k_+ k_- = 1$ for the inertial observer, and that $k_+ k_- = e^{-2 a \rho}$ 
for the uniformly accelerating observer. The time-translation vector 
field, defined in (\ref{eq:time}), is $\frac{\partial}{\partial \tau}$ 
for all observers. 

	Finally, note that $\grad^2 \tau = 0 = \grad^2 \rho$, and  
$\frac{\partial \tau}{\partial z} = \frac{\partial \rho}{\partial t}$ and 
$\frac{\partial \tau}{\partial t} = \frac{\partial \rho}{\partial z}$. Hence, 
if we Wick rotate $(t,z) \rightarrow (s \equiv i t, z)$ and $(\tau,\rho) 
\rightarrow (\xi \equiv i \tau, \rho)$ it follows that $\xi$, $\rho$ are harmonic, and 
$\xi + i \rho$ is an analytic (conformal) function of $s + i z$.

\subsection{Arbitrary 1+1D Spacetime}

	All 1+1 dimensional spacetimes can be written in the form:
$$ d s^2 = \Omega^2(z_+,z_-) d z_+ d z_- = \Omega^2 (d t^2 - d z^2) $$
	where $z_{\pm} = t \pm z$ as before. Much of the previous 
subsection carries over, with the main difference being the 
parametrisation of proper time, which must now satisfy $d \tau^2 = 
\Omega^2 d z_+ d z_-$ on the curve.
	
	Define:
\begin{align} k_-(z_-) & \equiv \Omega(z^{(\gamma)}_+(z_-),z_-) 
\sqrt{\frac{d z_+^{(\gamma)}(z_-)}{d z_-}} \notag \\
k_+(z_+) & \equiv \Omega(z_+,z^{(\gamma)}_-(z_+)) 
\sqrt{\frac{d z_-^{(\gamma)}(z_+)}{d z_+}} = 
\frac{\Omega(z_+,z^{(\gamma)}_-(z_+))^2}{k_-(z_-^{(\gamma)}(z_+))} \notag \\
 \tau^+(z_+) & \equiv \int_{z_0}^{z_+} k_+(u) d u \hs{1} \tau^-(z_-) 
\equiv \int_{z_-^{(\gamma)}(z_0)}^{z_-} k_-(u) d u \notag \end{align}
	We have still chosen the origin of the $(z_+,z_-)$ 
coordinate system to lie on the observers trajectory, but we no longer 
require that it coincide with the point of zero proper time. The point 
$\tau=0$ now occurs at $(z_+,z_-) = (z_0,z_-^{(\gamma)}(z_0))$. This is for 
later convenience. We still have $\tau(z_+,z_-) = \half(\tau^+(z_+) + 
\tau^-(z_-))$ and  $\rho(z_+,z_-) = \half | \tau^+(z_+) + \tau^-(z_-) |$, 
and the metric becomes:
\begin{equation} d s^2 = \frac{\Omega^2(z_+,z_-)}{k_+(z_+) k_-(z_-)} 
(d \tau^2 - d \rho^2) \label{eq:105} \end{equation}

	The time-translation vector field is still $\frac{\partial}{\partial \tau}$, 
and the final comments of Section 5.1, about the analyticity of the Wick-rotated spacetime, still apply.

\subsection{FRW Universes in 1+1 Dimensions}

	Consider a comoving observer in an FRW universe of 
arbitrary scale factor $a(t)$, in 1+1 dimensions:
$$ d s^2 = d t^2 - a(t)^2 d z^2 = C(\eta)^2 (d \eta^2 - d z^2) $$
where $C(\eta(t)) = a(t)$ and: 
\begin{equation} \eta(t) = \eta_0 + \int_0^t \frac{d t'}{a(t')} 
\label{eq:etadef} \end{equation}

This is a special case of Section 5.2. A comoving observer at $z=0$ satisfies 
$z_+^{(\gamma)}(z_-) = z_-$ where $z_{\pm} \equiv \eta \pm z$, so that 
\begin{equation} k_{\pm}(z_{\pm}) = C(z_{\pm}) \mbox{ and } \tau^{\pm}(z_{\pm}) = \int_{\eta_0}^{z_{\pm}} C(u) d u \label{eq:FRW1} \end{equation}
	We have chosen $z_0 = \eta_0$, so that on the observer's 
trajectory $\tau^+ = \tau^- = t$ (which is the observer's proper time). From 
(\ref{eq:105}), the metric can be written as 
\begin{equation} d s^2 = \frac{C(\half(z_+ + z_-))^2}{C(z_+) C(z_-)} (d \tau^2 
- d \rho^2) \label{eq:badmet} \end{equation}
	In this case we can invert $\tau^{\pm}(z_{\pm})$ explicitly, to get 
\begin{equation} z_{\pm} = \eta(\tau^{\pm}) \label{eq:FRW2} \end{equation}
(with $\eta(u)$ defined as in (\ref{eq:etadef})). This allows us to rewrite 
(\ref{eq:badmet}) in terms of $\tau,\rho$ as:
\begin{equation} d s^2 = \frac{C(\half \eta(\tau^+) + \half 
\eta(\tau_-))^2}{a(\tau^+) a(\tau^-)} (d \tau^2 - d \rho^2) 
\label{eq:goodmet} \end{equation}
	where $\tau^{\pm} = \tau \pm \rho$. 

	To generate insight into these results, we consider 3 examples.

\subsubsection{deSitter Space}

	DeSitter space has $a(t) = e^{\lambda t}$, which gives $\eta(t) = 
\frac{-1}{\lambda} e^{- \lambda t}$ and $C(\eta) = \frac{-1}{\lambda 
\eta}$. Note that $\eta(t) < 0$ for all $t$, so the domain of $\tau(x)$ 
will cover only the causal past of the point $(\eta,z) = (0,0)$, with 
particle horizons at $z = \mp \eta = \frac{\pm 1}{\lambda} 
e^{- \lambda t}$. From (\ref{eq:FRW1}) or (\ref{eq:FRW2}) we 
have $\tau^{\pm}(z_{\pm}) = \frac{-1}{\lambda} \log(-\lambda z_{\pm})$, so 
that $\tau,\rho$ are given by:

\begin{align}
\tau & = \frac{-1}{2 \lambda} \log(\lambda^2 z_+ z_-) = \frac{-1}{2 \lambda} 
\log(\lambda^2 (\eta^2 - z^2)) \notag \\
& = \frac{-1}{2 \lambda} \log(e^{-2 \lambda t} - \lambda^2 z^2) 
\label{eq:taudeSitt} \\
\rho & = | \frac{1}{2 \lambda} \log(\frac{e^{-\lambda t} + 
\lambda z}{e^{-\lambda t} - \lambda z})| = \frac{1}{2 \lambda} 
\log(\frac{e^{-\lambda t} + \lambda |z|}{e^{-\lambda t} - \lambda |z|}) 
\label{eq:rhodeSitt} \end{align}
	
\begin{figure}[h]
\center{\epsfig{figure=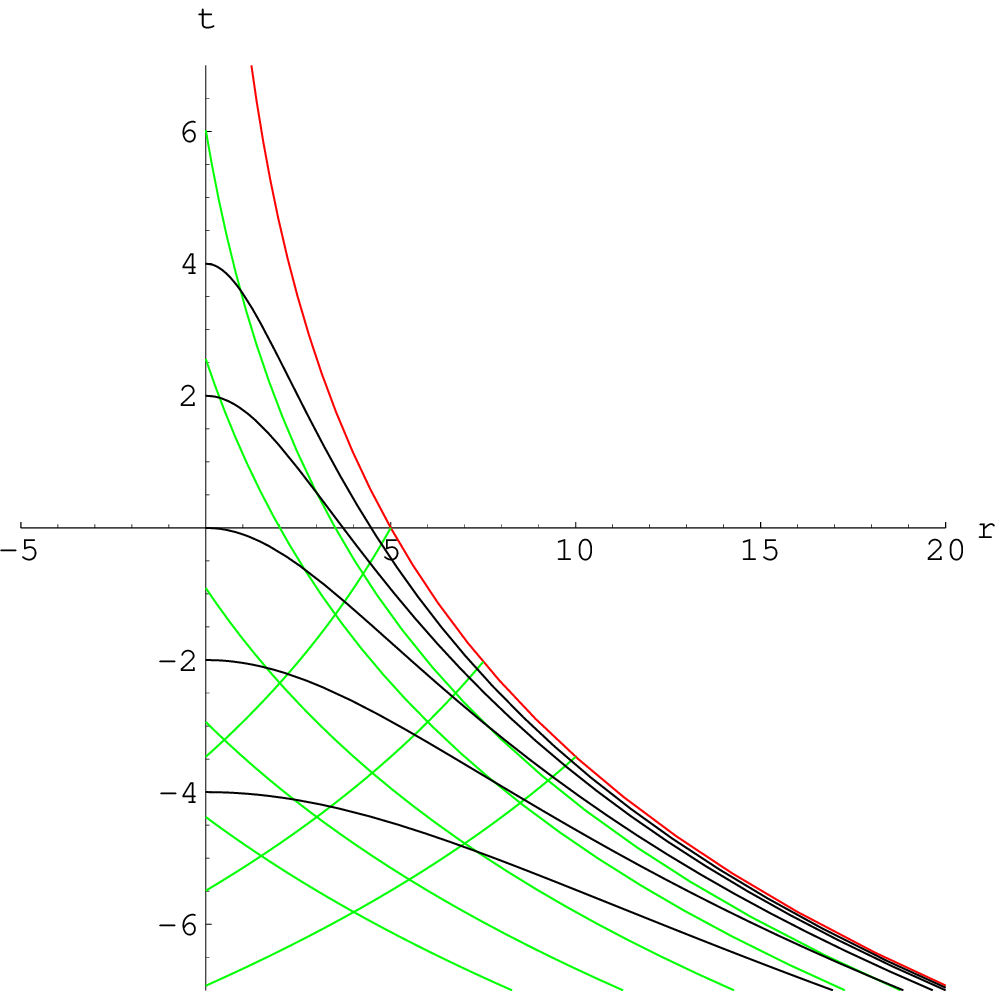, width=6cm}}
\caption{{\footnotesize DeSitter space in $(t,|z|)$ coordinates. The grey 
lines represent ingoing and outgoing photon trajectories (the outermost one 
is the particle horizon $z(t) = \frac{\pm e^{-\lambda t}}{\lambda}$), and  
the black lines are the observer's hypersurfaces of simultaneity for 
various values of $\tau_0$.}}
\end{figure}

The metric in $(\tau,\rho)$ coordinates, calculated either from (\ref{eq:goodmet}) or 
directly from 
(\ref{eq:taudeSitt}) and (\ref{eq:rhodeSitt}), is $d s^2 = \cosh^{-2}(\lambda \rho) 
(d \tau^2 - d \rho^2 )$. The time-translation vector field is given by:
\begin{equation} k = \frac{\partial}{\partial t} - \lambda z 
\frac{\partial}{\partial z} \hs{1} \mbox{or } = \frac{\partial}{\partial 
\tau} \mbox{ in } (\tau,\rho) \mbox{ coordinates} \notag \end{equation}
	Again this is a timelike Killing vector field on the domain of 
$\tau(x)$, and is spacelike outside this region. This $\tau$ 
coordinate, and the corresponding Killing vector field, are the same 
as those used in Gibbons and Hawkings original derivation of the 
thermal deSitter spectrum in 1977 \cite{GHaw}, and as in 
various other derivations since \cite{Lap,LoPa} (see also 
\cite{BD} and the references therein). The metric takes a more 
familiar form \cite{Schrod2} if we substitute $u = \frac{\tanh(\lambda \rho)}{\lambda}$, 
giving $d s^2 = (1 - \lambda^2 u^2) d \tau^2 - \frac{d u^2}{1 - 
\lambda^2 u^2}$. Hypersurfaces of 
constant $\tau$ have been plotted in Figure 9.

\subsubsection{The Milne Universe}

	The 1+1 D Milne universe is given by $d s^2 = d t^2 - a_0^2 t^2 d z^2$. 
Application of (\ref{eq:FRW1}) now gives $\tau^{\pm} = t e^{\pm a_0 z}$, so that:
\begin{equation} \tau(x) = t {\rm cosh}(a_0 z) \hs{.5} \rho(x) = 
|t {\rm sinh}(a_0 z)| \notag \end{equation}
	In terms of which the metric is $d s^2 = d \tau^2 - d \rho^2$. It 
is convenient to drop the absolute value in $\rho$ and to treat $\rho$ 
as a spatial (rather than a radial) coordinate. Although the 
cases $t>0$ and $t<0$ must be treated separately, a similar transformation 
holds in each case. With the coordinate singularity at 
$t=0$ moved to the light cone through the origin, the regions $t >0$ and 
$t<0$ are revealed to be regions F and P of Rindler space, and $\tau,\rho$ 
are the inertial coordinates of the underlying flat space. This is illustrated 
in Figure 10.

\begin{figure}[h]
\center{\epsfig{figure=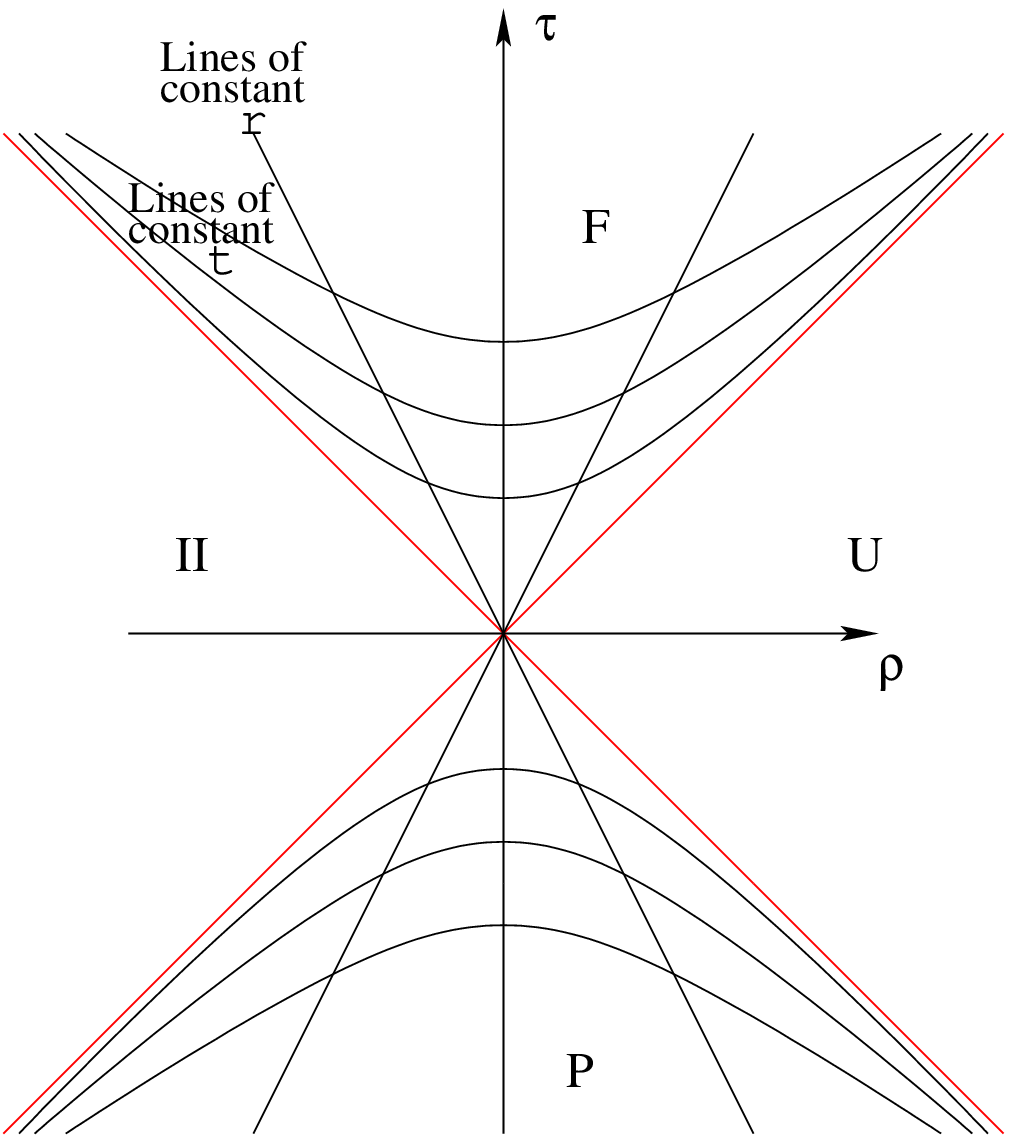, width=6cm}}
\caption{{\footnotesize The Milne universe in $(\tau,\rho)$ coordinates. The 
$(t,r)$ coordinate system corresponds to regions F and P of Section 5, 
but do not correspond to the radar coordinates of any observer. The 
radar coordinates $(\tau,\rho)$ of the `comoving' (inertial) observer 
recover the underlying flat space.}}
\end{figure}

	Since $t=0$ is just a coordinate singularity, there is no 
reason for the observer's trajectory to terminate there; the `completed' 
geodesic will cover all $t$. Clearly, as this observer charts her radar 
coordinates (sending and receiving light from various spacetime points)
it will be readily apparent to her that regions II and U also exist, and 
the domain of $(\tau,\rho)$ will certainly include these regions. The lines 
$\tau = \pm \rho$ which bound the original $(t,z)$ coordinate system are 
of no significance in the $(\tau,\rho)$ coordinate system.

	Some approaches to particle creation in the Milne 
universe (see \cite{BD} Section 5.3, and references therein) exploit 
the similarity with the Rindler case to derive non-trivial quantum 
effects in this spacetime. However, an important distinction must be made. 
Region U of Rindler space, expressed in Rindler 
coordinates, corresponds to spacetime as seen by a uniformly 
accelerating observer in flat space; it is this fact that gives 
physical meaning to the Rindler coordinate system. In the Milne universe, 
however, {\it no} observer would have $t$ as their radar 
time. A comoving observer at the origin of the Milne universe corresponds 
simply to an inertial observer (stationary at the origin) in flat space. It 
is therefore reassuring that the radar time of that observer brings  
us back to Minkowski space, and to the standard inertial vacuum.

\subsubsection{Radiation domination}

	In this case $a(t) = \sqrt{\lambda t}$, which gives:
\begin{align} \tau(x) & = t + \frac{\lambda}{4} z^2 \hs{1} \rho(x) = |z| \sqrt{\lambda t} 
\notag \\
& d s^2 = \half (1 + \frac{\tau}{\sqrt{\tau^2 - \rho^2}})(d \tau^2 - d \rho^2)
 \notag \end{align}
	Hypersurfaces of constant radar time are plotted in Figure 11.

\begin{figure}[h]
\center{\epsfig{figure=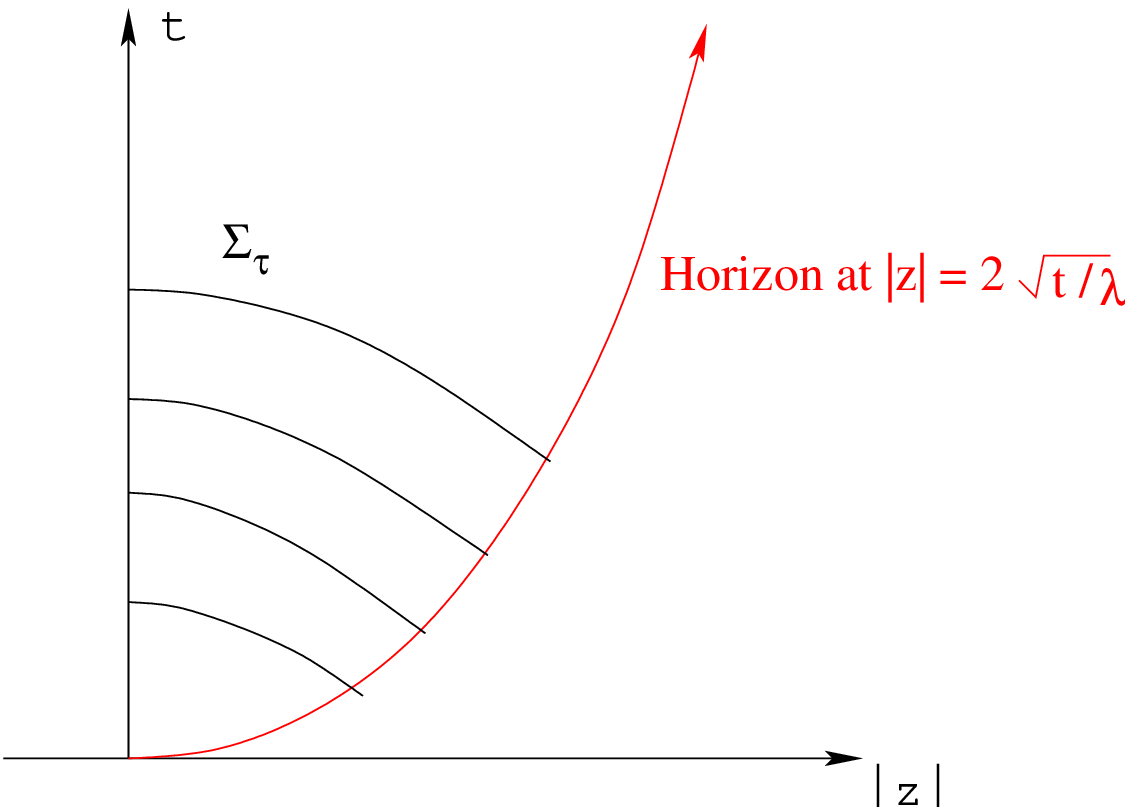, width=9cm}}
\caption{{\footnotesize Hypersurfaces of constant $\tau$ in an FRW universe 
with scale factor $a(t) = \sqrt{\lambda t}$, in $(t,|z|)$ coordinates. The 
domain of radar time covers only the causal future of the origin.}}
\end{figure}

	As a result of the singularity at $t=0$, the comoving observer's 
geodesic starts at finite proper time (chosen for 
convenience to be zero). The horizon now represents the future 
light cone $\rho = \tau$ (or $r = 2 \sqrt{\frac{t}{\lambda}}$) of the 
origin. This is not an acceleration horizon, and (unlike the Rindler case) 
the time translation vector field does not tend to zero there. Instead we find 
that, though the direction of the vector field $k^{\mu}(x)$ becomes null 
on the horizon, its components become infinite in this limit, so that $k^2 
\rightarrow \infty$ as the horizon is approached (i.e. as $\rho \rightarrow \tau$).

\subsection{3+1 D Cosmologies}

Consider the metric 
$$ d s^2 = d t^2 - a(t)^2 \left(d r^2 + f(r)^2 (d \theta^2 - 
\sin^2(\theta) d \phi^2)\right) $$ 
	(where $f(r) = \sin(r), r$ or $\rm{sinh}(r)$ for spatial sections 
which are respectively hyperbolic, flat, or closed), and consider a 
comoving observer at the origin. 
Only the radial null geodesics are relevant for determining $\tau^{\pm}$ for this observer, 
so that we can define $z^{\pm} = \eta \pm r$ and the 1+1 D results carry over 
almost unaltered. The $\tau^{\pm}$ are still given by (\ref{eq:FRW1}) or by 
inverting (\ref{eq:FRW2}), and the metric is given by:

\begin{equation} d s^2 = C(\half \eta(\tau^+) + \half \eta(\tau^-))^2 
\left(\frac{d \tau^2 - d \rho^2}{a(\tau^+) a(\tau^-)} + 
f(r(\tau^+,\tau^-))^2 (d \theta^2 - \sin^2(\theta) d \phi^2)\right) 
 \notag \end{equation}
	where $r(\tau^+,\tau^-) = \half (\eta(\tau^+) - \eta(\tau^-))$. The 
examples of Section 5.3 extend almost unaltered to the 3+1 dimensional case.

\section{DISCUSSION}

We have presented a formulation of fermionic quantum field
theory in electromagnetic and gravitational backgrounds that is analogous to 
the methods used in multiparticle quantum 
mechanics. The main difference is in the particle interpretation, which 
requires us to consider the entire Dirac Sea, as well as any particles 
which may be present. This approach provides a
conceptually transparent approach to the theory and a  
simple derivation of the general S-Matrix element and expectation value of the theory. Moreover, it also leads to a consistent particle interpretation 
for all times and any background, without 
requiring any `asymptotic niceness conditions' on the `in' and `out' states. Other
advantages include the ease with which unitarity of the S-Matrix follows 
from conservation of the Dirac inner product,
insights into quantum anomalies, and the fact 
that Hermitian extension provides well-defined second quantized operators without 
requiring a complete set of orthonormal modes. We have used the 
concept of `radar time' to  
generalise the particle interpretation to an arbitrarily moving observer, providing a 
definition of particle which depends only on the observer's motion and on the 
background, and not on the choice of coordinates, the choice of gauge, or the 
detailed construction of the particle detector.

Ever since the pioneering work of Unruh \cite{Unruh} and Davies 
\cite{Davies} in 1975, it has been known that the 
concept of particle differs for different observers. However, attempts to 
systematically assign a choice of particle/antiparticle modes uniquely to a 
given observer have not been successful. Such definitions have either relied on 
the existence of suitable symmetries (Killing vectors, conformal symmetry etc) or 
on an arbitrary choice of foliation of spacetime into `space' and `time' (see for 
instance \cite{BD,Full} and references therein). A notable exception  
is reference \cite{ZAZ} which treated arbitrary observers in flat space, using a 
foliation defined uniquely in terms of the motion of the observer. However, as 
well as being only applicable in flat space, their foliation is often multivalued, and  
omits portions of the observers causal envelope if discrete changes in velocity are allowed \cite{Me2}. 

Though particle detector models provide a useful operational particle concept, it 
is inherently circular for a particle detector to be anything 
that detects particles, and a particle to be `anything 
detected by a particle detector'. It is also known \cite{Sr1,Sr2} that in 
electromagnetic backgrounds (and in cases where particles could already be defined independtly of detector models) the predictions of particle detector models 
are not always proportional to the number of particles present, even when 
the detector is inertial. We might conclude that a particle concept 
is only an approximate notion, or even that `particles do not exist' 
\cite{Dav}. However, the observer-dependent particle 
interpretation presented here (and in \cite{Me1}) 
averts this pessimistic conclusion, and provides a concrete answer to 
to the question ``what do particle detectors detect?'' 

We hope that the present work, along with \cite{Me1,mythesis}, has shown the computational and
conceptual value of working with a concrete representation of 
the Dirac Sea. We strongly support  Jackiw's claim~\cite{Jackiw} that 
``physical consequences can be drawn from Dirac's construction''. Care has been taken to ensure that the dynamics is kept separate from 
the kinematics. The evolution equation is explicitly local and causal. Though the   
categorisation of states in terms of their particle content requires an observer-dependent 
foliation of spacetime, this in no way affects the evolution of these states. This 
goes some way towards showing that the foliation dependence of quantum mechanics 
need not conflict with the coordinate covariance of general relativity, 
provided one remembers the important 
role played in both theories by the observer.

\appendix{Green Functions and Projection Operators}

In equations (\ref{eq:fluc11}) - (\ref{eq:fluc12}) the vacuum 
expectation value of a 
general `one-particle' operator $\hat{A}_{\rm{phys}}$ (the physical 
extension of some operator $\hat{A}_1:\clh \rightarrow \clh$) was expressed in terms 
of traces 
of projection operators $\hat{P}^{\pm}_{|F\ra}(\tau)$ and 
$\hat{P}^{\pm}_{\tau}$. It is common in many textbooks \cite{BD,Full,GMR}
to express expectation values in terms of traces of (2-point) Greens  
functions. This appendix summarises the connections between 
projection operators and Greens functions. We 
also make connection with the ``first order density matrix'' or 
``Dirac density matrix'' of multiparticle quantum mechanics (see 
\cite{MYS} pgs 8-10 for instance), emphasising the role of the 
negative-energy Wightman function as the Dirac density matrix 
of the Dirac Sea. The observer-dependent particle interpretation 
presented in Section 3, along with the corresponding time-dependent 
vacuum, allows the definition of a time-dependent family of 
`vacuum Greens functions'. By presenting these as special cases of general 
state-dependent Greens functions, further clarification can be made of the 
connection between the negative-energy Wightman function and the 
Dirac density matrix.

In the quantum mechanics of non-relativistic fermions, a common 
tool (see e.g. \cite{MYS}, pg 8-10) is the {\it first order density 
matrix} $\gamma(\br',\sigma',\br,\sigma)$, defined from the normalised 
many-body wave function $\Phi(\br_1,\sigma_1, \dots ,\br_N,\sigma_N)$ by:

$$ \gamma_{\Phi}(\br',\sigma',\br,\sigma) = N \sum_{\sigma_2,\dots,\sigma_N} 
\int {\rm d} \br_2\dots {\rm d} \br_N \Phi(\br,\sigma, \dots ,\br_N,\sigma_N) 
\Phi^*(\br',\sigma', \dots ,\br_N,\sigma_N)$$
When the wavefunction is a Slater determinant of one-particle states 
$\phi_i$, this takes the simpler form:
$$ \gamma_{\Phi}(\br',\sigma',\br,\sigma) = \sum_{i=1}^{N} 
\phi_i(\br,\sigma) \phi_i^*(\br',\sigma') $$
	and is referred to as the {\it Dirac density matrix}. Consider 
now the relativistic case, and take the 
multiparticle wavefunction to be the `evolved vacuum' 
$|{\rm vac}_{\tau_0}(\tau)\ra$ from (\ref{eq:forapp1}). Then the Dirac 
density matrix `at time $\tau$' is:
$$ \gamma_{|{\rm vac}_{\tau_0}(\tau)\ra}(y|_{\tau},x|_{\tau}) = \sum_i v_{i,\tau_0}(x|_{\tau}) 
\bar{v}_{i,\tau_0}(y|_{\tau}) $$
	(which is a $4 \times 4$ matrix, implicitly containing 
the spin dependence). From (\ref{eq:p-}) this is simply the negative 
energy projection operator $\hat{P}^-_{\tau_0}(\tau)$ in  
coordinate representation. More precisely:

$$\hat{P}^-_{\tau_0}(\tau) \psi(x|_{\tau}) = \int_{\Sigma_\tau} e(y) 
\gamma_{|{\rm vac}_{\tau_0}(\tau)\ra}(y|_{\tau},x|_{\tau}) 
\gamma^{\mu}(y) \psi(y|_{\tau}) d \Sigma_{\mu}(y) $$
The connection between $\hat{P}^-_{\tau_0}(\tau)$, $\gamma_{|{\rm vac}_{\tau_0}(\tau)\ra}(y,x)$ and the {\it negative energy Wightman function} 
$S^-(x,y) \equiv \la 0 | \bar{\hat{\psi}}(y) \hat{\psi}(x) | 0 \ra$  
also follows. To see this, introduce the Schr\"{o}dinger 
picture field operator $\hat{\psi}(x|_{\Sigma}) \equiv \sum_i 
\psi_i(x|_{\Sigma}) i_{\psi_i(x|_{\Sigma})}$, as in 
multiparticle quantum mechanics. The connection with Canonical methods is
straightforward \cite{Me1}. The field operator can be written in the 
Heisenberg picture as $\hat{\psi}(x) = \sum_i 
\psi_i(x) i_{\psi_i(x|_{\Sigma_r})}$, where $\Sigma_r$ is an 
arbitrary fixed `reference hypersurface' on which the Heisenberg 
picture states are defined. If we choose the `vacuum' $|0 \ra$ to be 
$|\rm{vac}_{\tau_0}(\Sigma_r)\ra$ (which is the Heisenberg picture representative of $| \rm{vac}_{\tau_0}(\tau)\ra$) then we 
have
$$ S_{\tau_0}^-(x,y) = \sum_i v_{i,\tau_0}(x) 
\bar{v}_{i,\tau_0}(y) $$
	where the subscript $\tau_0$ signifies that 
$S_{\tau_0}^-(x,y)$ depends on $\tau_0$ through the 
choice of vacuum state $|{\rm vac}_{\tau_0}(\tau)\ra$. Therefore 
$\gamma_{|{\rm vac}_{\tau_0}(\tau)\ra}(y|_{\tau},x|_{\tau}) 
= S_{\tau_0}^-(x|_{\tau},y_{\tau})$. That is, the negative energy Wightman 
function is the Dirac density function of the Dirac Sea, and is the kernel 
of $\hat{P}^-_{\tau_0}(\tau)$. The {\it positive energy Wightman function} 
$S_{\tau_0}^+(x,y)$ can similarly be written as:
$$S_{\tau_0}^+(x,y) \equiv \ra \rm{vac}_{\tau_0}(\Sigma_r)|\hat{\psi}(x) 
\bar{\hat{\psi}}(y) |\rm{vac}_{\tau_0}(\Sigma_r)\ra = \sum_i u_{i,\tau_0}(x) 
\bar{u}_{i,\tau_0}(y)$$
which is the kernel of $\hat{P}^+_{\tau_0}(\tau)$.

	In terms of $S_{\tau_0}^{\pm}(x,y)$, other commonly used 2-point 
functions \cite{Full} are the: 
\[ \begin{array}{lrl}
\mbox{{\it Feynman propogator}} & i S_{F,\tau_0}(x,y) & = 
\theta(\tau_x - \tau_y) S^+_{\tau_0}(x,y) - \theta(\tau_y - \tau_x) 
S^-_{\tau_0}(x,y) \notag \\
\mbox{{\it Hadamard function}} & S^{(1)}_{\tau_0}(x,y) 
 & = S^+_{\tau_0}(x,y) - S^-_{\tau_0}(x,y) \notag \\
\mbox{{\it full propogator}} & i S(x,y) 
& = S^+_{\tau_0}(x,y) + S^-_{\tau_0}(x,y) \notag \\
\mbox{{\it retarded propogator}} & S_{R}(x,y) & = - 
\theta(\tau_x - \tau_y) S(x,y) \notag \\
\mbox{{\it advanced propogator}} & S_{A}(x,y) 
& = \theta(\tau_y - \tau_x) S(x,y) \notag \end{array} \]

$S^{(1)}_{\tau_0}(x|_{\tau},y|_{\tau})$ is the kernel of the `first 
quantized' number operator $\hat{N}_{1,\tau_0}(\tau)$, while 
$i S(x|_{\tau},y|_{\tau'})$ is the kernel of the `first 
quantized' evolution operator $\hat{U}_{1}(\tau,\tau')$. $S(x,y)$, 
$S_{R}(x,y)$ and $S_{A}(x,y)$ are independent of $\tau_0$, since they 
do not depend on any choice of state. Furthermore they do not depend on the 
decomposition of solutions into positive/negative frequency modes.

	The Wightman function $S_{\tau_0}^{-}(x,y)$ can be generalised to 
an arbitrary state $|F(\Sigma_r)\ra$ (which is the Heisenberg 
picture representative of $|F(\tau)\ra$) as:

\begin{align} S^-_{|F\ra}(x,y) & \equiv \la F(\Sigma_r)| \bar{\hat{\psi}}(y) \hat{\psi}(x) | F(\Sigma_r) \ra \notag \\
 & = \sum_{i \in I} \psi_i(x) \bar{\psi}(y) = \gamma_{|F\ra}(y,x) \notag \end{align}

Again, the negative energy Wightman function is just the Dirac 
density matrix of the state in question, and is the kernel of the 
operator $\hat{P}^-_{|F\ra}(\tau)$ defined in Section 3.3. Here $S^+_{|F\ra}(x,y)$ is the kernel of $\hat{P}^+_{|F\ra}(\tau)$.

\begin{acknowledgments}

	 We thank Anton Garrett for helpful 
discussions. Carl Dolby also thanks Merton College, Oxford for financial support.

\end{acknowledgments}

\end{article}

\end{document}